\documentclass[12pt,a4paper]{article}
\usepackage{amsmath}
\usepackage{xspace}
\usepackage{epsfig}
\usepackage{rotating}
\usepackage{dcolumn}
\usepackage{url}
\usepackage{hyperref}
\usepackage[numbers,sort&compress]{natbib}
\usepackage{hypernat}
\hypersetup{ pdfborder=0 0 0, urlbordercolor=1 0 0 }

\newcommand{\citenumfont}{}
\usepackage{graphicx}
\usepackage{color}
\usepackage{colortbl}
\begin{document}

\newcommand{\Kshort}{\ensuremath{K^0_{\rm S}}}


\begin{titlepage}
\belowpdfbookmark{Title page}{title}

\pagenumbering{roman}
\vspace*{-1.5cm}
\centerline{\large EUROPEAN ORGANIZATION FOR NUCLEAR RESEARCH (CERN)}
\vspace*{1.5cm}
\hspace*{-5mm}\begin{tabular*}{16cm}{lc@{\extracolsep{\fill}}r}
\vspace*{-12mm}\mbox{\!\!\!\epsfig{figure=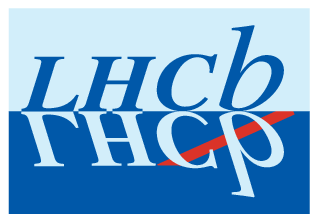,width=.12\textwidth}}& & \\
&& CERN-PH-EP-2010-027 \\
&& 18 August 2010, rev.\ 15 September 2010 \\
\end{tabular*}
\vspace*{4cm}
\begin{center}
\Large
{\bf \boldmath
\huge {\boldmath Prompt \Kshort\ production in\\ $pp$ collisions at $\sqrt{s}= 0.9$~TeV}\\
\vspace*{2cm}
 }
\normalsize {
The LHCb Collaboration%
\footnote{Authors are listed on the following pages.}
}
\end{center}
\vspace{\fill}
\centerline{\bf Abstract}
\vspace*{5mm}\noindent 
The production of \Kshort\ mesons in $pp$ collisions 
at a centre-of-mass energy 
of  0.9~TeV is studied with the LHCb detector 
at the Large Hadron Collider. 
The luminosity of the analysed sample is determined using a novel technique,
involving measurements of the beam currents, sizes and positions, 
and is found to be $6.8\pm1.0~{\rm \mu b}^{-1}$. 
The differential prompt \Kshort\ production cross-section is measured
as a function of the \Kshort\ transverse momentum and rapidity in
the region $0 < p_{\rm T} < 1.6~{\rm GeV}/c$ and $2.5 < y < 4.0$. 
The data are found to be in reasonable agreement with previous measurements
and generator expectations.

\vspace*{2.cm}\noindent
{\em Keywords:} strangeness, production, cross-section, luminosity, LHC, LHCb
\vspace{\fill}

\end{titlepage}
\newpage
\setcounter{page}{2}
\mbox{~}
\newpage

\belowpdfbookmark{LHCb author list}{authors}
\centerline{\large\bf The LHCb Collaboration}
\begin{flushleft}
\small
R.~Aaij$^{23}$, 
C.~Abellan~Beteta$^{35,m}$, 
B.~Adeva$^{36}$, 
M.~Adinolfi$^{42}$, 
C.~Adrover$^{6}$, 
A.~Affolder$^{48}$, 
M.~Agari$^{10}$, 
Z.~Ajaltouni$^{5}$, 
J.~Albrecht$^{37}$, 
F.~Alessio$^{6,37}$, 
M.~Alexander$^{47}$, 
M.~Alfonsi$^{18}$, 
P.~Alvarez~Cartelle$^{36}$, 
A.A.~Alves~Jr$^{22}$, 
S.~Amato$^{2}$, 
Y.~Amhis$^{38}$, 
J.~Amoraal$^{23}$, 
J.~Anderson$^{39}$, 
R.~Antunes~Nobrega$^{22,k}$, 
R.~Appleby$^{50}$, 
O.~Aquines~Gutierrez$^{10}$, 
A.~Arefyev$^{30}$, 
L.~Arrabito$^{53}$, 
M.~Artuso$^{52}$, 
E.~Aslanides$^{6}$, 
G.~Auriemma$^{22,l}$, 
S.~Bachmann$^{11}$, 
Y.~Bagaturia$^{11}$, 
D.S.~Bailey$^{50}$, 
V.~Balagura$^{30,37}$, 
W.~Baldini$^{16}$, 
G.~Barber$^{49}$, 
C.~Barham$^{43}$, 
R.J.~Barlow$^{50}$, 
S.~Barsuk$^{7}$,
S.~Basiladze$^{31}$, 
A.~Bates$^{47}$, 
C.~Bauer$^{10}$, 
Th.~Bauer$^{23}$, 
A.~Bay$^{38}$, 
I.~Bediaga$^{1}$, 
T.~Bellunato$^{20,i}$, 
K.~Belous$^{34}$, 
I.~Belyaev$^{23,30}$, 
M.~Benayoun$^{8}$, 
G.~Bencivenni$^{18}$, 
R.~Bernet$^{39}$, 
R.P.~Bernhard$^{39}$, 
M.-O.~Bettler$^{17}$, 
M.~van~Beuzekom$^{23}$, 
J.H.~Bibby$^{51}$, 
S.~Bifani$^{12}$, 
A.~Bizzeti$^{17,g}$, 
P.M.~Bj\o rnstad$^{50}$, 
T.~Blake$^{49}$, 
F.~Blanc$^{38}$, 
C.~Blanks$^{49}$, 
J.~Blouw$^{11}$, 
S.~Blusk$^{52}$, 
A.~Bobrov$^{33}$, 
V.~Bocci$^{22}$, 
B.~Bochin$^{29}$, 
E.~Bonaccorsi$^{37}$, 
A.~Bondar$^{33}$, 
N.~Bondar$^{29,37}$, 
W.~Bonivento$^{15}$, 
S.~Borghi$^{47}$, 
A.~Borgia$^{52}$, 
E.~Bos$^{23}$, 
T.J.V.~Bowcock$^{48}$, 
C.~Bozzi$^{16}$, 
T.~Brambach$^{9}$, 
J.~van~den~Brand$^{24}$, 
L.~Brarda$^{37}$, 
J.~Bressieux$^{38}$, 
S.~Brisbane$^{51}$, 
M.~Britsch$^{10}$, 
N.H.~Brook$^{42}$, 
H.~Brown$^{48}$, 
S.~Brusa$^{16}$, 
A.~B\"uchler-Germann$^{39}$, 
A.~Bursche$^{39}$, 
J.~Buytaert$^{37}$, 
S.~Cadeddu$^{15}$, 
J.M.~Caicedo~Carvajal$^{37}$, 
O.~Callot$^{7}$, 
M.~Calvi$^{20,i}$, 
M.~Calvo~Gomez$^{35,m}$, 
A.~Camboni$^{35}$, 
W.~Cameron$^{49}$, 
L.~Camilleri$^{37}$, 
P.~Campana$^{18}$, 
A.~Carbone$^{14}$, 
G.~Carboni$^{21,j}$, 
R.~Cardinale$^{19,h}$, 
A.~Cardini$^{15}$, 
J.~Carroll$^{48}$, 
L.~Carson$^{36}$, 
K.~Carvalho~Akiba$^{23}$, 
G.~Casse$^{48}$, 
M.~Cattaneo$^{37}$, 
B.~Chadaj$^{37}$, 
M.~Charles$^{51}$, 
Ph.~Charpentier$^{37}$, 
J.~Cheng$^{3}$, 
N.~Chiapolini$^{39}$, 
A.~Chlopik$^{27}$, 
J.~Christiansen$^{37}$, 
P.~Ciambrone$^{18}$, 
X.~Cid~Vidal$^{36}$, 
P.J.~Clark$^{46}$, 
P.E.L.~Clarke$^{46}$, 
M.~Clemencic$^{37}$, 
H.V.~Cliff$^{43}$, 
J.~Closier$^{37}$, 
C.~Coca$^{28}$, 
V.~Coco$^{52}$, 
J.~Cogan$^{6}$, 
P.~Collins$^{37}$, 
A.~Comerma-Montells$^{35}$, 
F.~Constantin$^{28}$, 
G.~Conti$^{38}$, 
A.~Contu$^{51}$, 
P.~Cooke$^{48}$, 
M.~Coombes$^{42}$, 
B.~Corajod$^{37}$, 
G.~Corti$^{37}$, 
G.A.~Cowan$^{46}$, 
R.~Currie$^{46}$, 
B.~D'Almagne$^{7}$, 
C.~D'Ambrosio$^{37}$, 
I.~D'Antone$^{14}$, 
W.~Da~Silva$^{8}$, 
E.~Dane'$^{18}$, 
P.~David$^{8}$, 
I.~De~Bonis$^{4}$, 
S.~De~Capua$^{21,j}$, 
M.~De~Cian$^{39}$, 
F.~De~Lorenzi$^{12}$, 
J.M.~De~Miranda$^{1}$, 
L.~De~Paula$^{2}$, 
P.~De~Simone$^{18}$, 
D.~Decamp$^{4}$, 
G.~Decreuse$^{37}$, 
H.~Degaudenzi$^{38,37}$, 
M.~Deissenroth$^{11}$, 
L.~Del~Buono$^{8}$, 
C.J.~Densham$^{45}$, 
C.~Deplano$^{15}$, 
O.~Deschamps$^{5}$, 
F.~Dettori$^{15,c}$, 
J.~Dickens$^{43}$, 
H.~Dijkstra$^{37}$, 
M.~Dima$^{28}$, 
S.~Donleavy$^{48}$, 
P.~Dornan$^{49}$, 
D.~Dossett$^{44}$, 
A.~Dovbnya$^{40}$, 
R.~Dumps$^{37}$, 
F.~Dupertuis$^{38}$, 
L.~Dwyer$^{48}$, 
R.~Dzhelyadin$^{34}$, 
C.~Eames$^{49}$, 
S.~Easo$^{45}$, 
U.~Egede$^{49}$, 
V.~Egorychev$^{30}$, 
S.~Eidelman$^{33}$, 
D.~van~Eijk$^{23}$, 
F.~Eisele$^{11}$, 
S.~Eisenhardt$^{46}$, 
L.~Eklund$^{47}$, 
D.G.~d'Enterria$^{35,n}$, 
D.~Esperante~Pereira$^{36}$, 
L.~Est\`eve$^{43}$, 
E.~Fanchini$^{20,i}$, 
C.~F\"arber$^{11}$, 
G.~Fardell$^{46}$, 
C.~Farinelli$^{23}$, 
S.~Farry$^{12}$, 
V.~Fave$^{38}$, 
G.~Felici$^{18}$, 
V.~Fernandez~Albor$^{36}$, 
M.~Ferro-Luzzi$^{37}$, 
S.~Filippov$^{32}$, 
C.~Fitzpatrick$^{46}$, 
W.~Flegel$^{37}$,
F.~Fontanelli$^{19,h}$, 
C.~Forti$^{18}$, 
R.~Forty$^{37}$, 
C.~Fournier$^{37}$, 
B.~Franek$^{45}$, 
M.~Frank$^{37}$, 
C.~Frei$^{37}$, 
M.~Frosini$^{17,e}$, 
J.L.~Fungueirino~Pazos$^{36}$, 
S.~Furcas$^{20}$, 
A.~Gallas~Torreira$^{36}$, 
D.~Galli$^{14,b}$, 
M.~Gandelman$^{2}$, 
P.~Gandini$^{51}$, 
Y.~Gao$^{3}$, 
J-C.~Garnier$^{37}$, 
L.~Garrido$^{35}$, 
D.~Gascon$^{35}$, 
C.~Gaspar$^{37}$, 
A.~Gaspar~De~Valenzuela~Cue$^{35,m}$, 
J.~Gassner$^{39}$, 
N.~Gauvin$^{38}$, 
P.~Gavillet$^{37}$, 
M.~Gersabeck$^{37}$, 
T.~Gershon$^{44}$, 
Ph.~Ghez$^{4}$, 
V.~Gibson$^{43}$, 
Yu.~Gilitsky$^{34,\dagger}$, 
V.V.~Gligorov$^{37}$, 
C.~G\"obel$^{54}$, 
D.~Golubkov$^{30}$, 
A.~Golutvin$^{49,30,37}$, 
A.~Gomes$^{1}$, 
G.~Gong$^{3}$, 
H.~Gong$^{3}$, 
H.~Gordon$^{51}$, 
M.~Grabalosa~G\'andara$^{35}$, 
V.~Gracco$^{19,h}$, 
R.~Graciani~Diaz$^{35}$, 
L.A.~Granado~Cardoso$^{37}$, 
E.~Graug\'es$^{35}$, 
G.~Graziani$^{17}$, 
A.~Grecu$^{28}$, 
S.~Gregson$^{43}$, 
G.~Guerrer$^{1}$, 
B.~Gui$^{52}$, 
E.~Gushchin$^{32}$, 
Yu.~Guz$^{34,37}$, 
Z.~Guzik$^{27}$, 
T.~Gys$^{37}$, 
G.~Haefeli$^{38}$, 
S.C.~Haines$^{43}$, 
T.~Hampson$^{42}$, 
S.~Hansmann-Menzemer$^{11}$, 
R.~Harji$^{49}$, 
N.~Harnew$^{51}$, 
P.F.~Harrison$^{44}$, 
J.~He$^{7}$, 
K.~Hennessy$^{48}$, 
P.~Henrard$^{5}$, 
J.A.~Hernando~Morata$^{36}$, 
E.~van~Herwijnen$^{37}$, 
A.~Hicheur$^{38}$, 
E.~Hicks$^{48}$, 
H.J.~Hilke$^{37}$, 
W.~Hofmann$^{10}$, 
K.~Holubyev$^{11}$, 
P.~Hopchev$^{4}$, 
W.~Hulsbergen$^{23}$, 
P.~Hunt$^{51}$, 
T.~Huse$^{48}$, 
R.S.~Huston$^{12}$, 
D.~Hutchcroft$^{48}$, 
F.~Iacoangeli$^{22}$, 
V.~Iakovenko$^{7,41}$, 
C.~Iglesias~Escudero$^{36}$, 
C.~Ilgner$^{9}$, 
J.~Imong$^{42}$, 
R.~Jacobsson$^{37}$, 
M.~Jahjah~Hussein$^{5}$, 
O.~Jamet$^{37}$, 
E.~Jans$^{23}$, 
F.~Jansen$^{23}$, 
P.~Jaton$^{38}$, 
B.~Jean-Marie$^{7}$, 
M.~John$^{51}$, 
D.~Johnson$^{51}$, 
C.R.~Jones$^{43}$, 
B.~Jost$^{37}$, 
F.~Kapusta$^{8}$, 
T.M.~Karbach$^{9}$, 
A.~Kashchuk$^{29}$, 
S.~Katvars$^{43}$, 
J.~Keaveney$^{12}$, 
U.~Kerzel$^{43}$, 
T.~Ketel$^{24}$, 
A.~Keune$^{38}$, 
S.~Khalil$^{52}$, 
B.~Khanji$^{6}$, 
Y.M.~Kim$^{46}$, 
M.~Knecht$^{38}$, 
S.~Koblitz$^{37}$, 
A.~Konoplyannikov$^{30}$, 
P.~Koppenburg$^{23}$, 
M.~Korolev$^{31}$, 
A.~Kozlinskiy$^{23}$, 
L.~Kravchuk$^{32}$, 
R.~Kristic$^{37}$, 
G.~Krocker$^{11}$, 
P.~Krokovny$^{11}$, 
F.~Kruse$^{9}$, 
K.~Kruzelecki$^{37}$, 
M.~Kucharczyk$^{25}$, 
I.~Kudryashov$^{31}$, 
S.~Kukulak$^{25}$, 
R.~Kumar$^{14}$, 
T.~Kvaratskheliya$^{30}$, 
V.N.~La~Thi$^{38}$, 
D.~Lacarrere$^{37}$, 
A.~Lai$^{15}$, 
R.W.~Lambert$^{37}$, 
G.~Lanfranchi$^{18}$, 
C.~Langenbruch$^{11}$, 
T.~Latham$^{44}$, 
R.~Le~Gac$^{6}$, 
J.-P.~Lees$^{4}$, 
R.~Lef\`evre$^{5}$, 
A.~Leflat$^{31,37}$, 
J.~Lefran\c cois$^{7}$, 
F.~Lehner$^{39}$, 
M.~Lenzi$^{17}$, 
O.~Leroy$^{6}$, 
T.~Lesiak$^{25}$, 
L.~Li$^{3}$, 
Y.Y.~Li$^{43}$, 
L.~Li~Gioi$^{5}$, 
J.~Libby$^{51}$, 
M.~Lieng$^{9}$, 
R.~Lindner$^{37}$, 
S.~Lindsey$^{48}$, 
C.~Linn$^{11}$, 
B.~Liu$^{3}$, 
G.~Liu$^{37}$, 
S.~L\"ochner$^{10}$, 
J.H.~Lopes$^{2}$, 
E.~Lopez~Asamar$^{35}$, 
N.~Lopez-March$^{38}$, 
P.~Loveridge$^{45}$, 
J.~Luisier$^{38}$, 
B.~M'charek$^{24}$, 
F.~Machefert$^{7}$, 
I.V.~Machikhiliyan$^{4,30}$, 
F.~Maciuc$^{10}$, 
O.~Maev$^{29}$, 
J.~Magnin$^{1}$, 
A.~Maier$^{37}$, 
S.~Malde$^{51}$, 
R.M.D.~Mamunur$^{37}$, 
G.~Manca$^{15,c}$, 
G.~Mancinelli$^{6}$, 
N.~Mangiafave$^{43}$, 
U.~Marconi$^{14}$, 
R.~M\"arki$^{38}$, 
J.~Marks$^{11}$, 
G.~Martellotti$^{22}$, 
A.~Martens$^{7}$, 
L.~Martin$^{51}$, 
D.~Martinez~Santos$^{36}$, 
A.~Massaferri$^{1}$, 
Z.~Mathe$^{12}$, 
C.~Matteuzzi$^{20}$, 
V.~Matveev$^{34}$, 
E.~Maurice$^{6}$, 
B.~Maynard$^{52}$, 
A.~Mazurov$^{32}$, 
G.~McGregor$^{50}$, 
R.~McNulty$^{12}$, 
C.~Mclean$^{14}$, 
M.~Merk$^{23}$, 
J.~Merkel$^{9}$, 
M.~Merkin$^{31}$, 
R.~Messi$^{21,j}$, 
F.C.D.~Metlica$^{42}$, 
S.~Miglioranzi$^{37}$, 
M.-N.~Minard$^{4}$, 
G.~Moine$^{37}$, 
S.~Monteil$^{5}$, 
D.~Moran$^{12}$, 
J.~Morant$^{37}$, 
J.V.~Morris$^{45}$, 
J.~Moscicki$^{37}$, 
R.~Mountain$^{52}$, 
I.~Mous$^{23}$, 
F.~Muheim$^{46}$, 
R.~Muresan$^{38}$, 
F.~Murtas$^{18}$, 
B.~Muryn$^{26}$, 
M.~Musy$^{35}$, 
J.~Mylroie-Smith$^{48}$, 
P.~Naik$^{42}$, 
T.~Nakada$^{38}$, 
R.~Nandakumar$^{45}$, 
J.~Nardulli$^{45}$, 
A.~Nawrot$^{27}$, 
M.~Nedos$^{9}$, 
M.~Needham$^{38}$, 
N.~Neufeld$^{37}$, 
P.~Neustroev$^{29}$, 
M.~Nicol$^{7}$, 
L.~Nicolas$^{38}$, 
S.~Nies$^{9}$, 
V.~Niess$^{5}$, 
N.~Nikitin$^{31}$, 
A.~Noor$^{48}$, 
A.~Oblakowska-Mucha$^{26}$, 
V.~Obraztsov$^{34}$, 
S.~Oggero$^{23}$, 
O.~Okhrimenko$^{41}$, 
R.~Oldeman$^{15,c}$, 
M.~Orlandea$^{28}$, 
A.~Ostankov$^{34}$, 
J.~Palacios$^{23}$, 
M.~Palutan$^{18}$, 
J.~Panman$^{37}$, 
A.~Papadelis$^{23}$, 
A.~Papanestis$^{45}$, 
M.~Pappagallo$^{13,a}$, 
C.~Parkes$^{47}$, 
C.J.~Parkinson$^{49}$, 
G.~Passaleva$^{17}$, 
G.D.~Patel$^{48}$, 
M.~Patel$^{49}$, 
S.K.~Paterson$^{49,37}$, 
G.N.~Patrick$^{45}$, 
C.~Patrignani$^{19,h}$, 
E.~Pauna$^{28}$, 
C.~Pauna~(Chiojdeanu)$^{28}$, 
C.~Pavel~(Nicorescu)$^{28}$, 
A.~Pazos~Alvarez$^{36}$, 
A.~Pellegrino$^{23}$, 
G.~Penso$^{22,k}$, 
M.~Pepe~Altarelli$^{37}$, 
S.~Perazzini$^{14,b}$, 
D.L.~Perego$^{20,i}$, 
E.~Perez~Trigo$^{36}$, 
A.~P\'erez-Calero~Yzquierdo$^{35}$, 
P.~Perret$^{5}$, 
G.~Pessina$^{20}$, 
A.~Petrella$^{16,d}$, 
A.~Petrolini$^{19,h}$, 
E.~Picatoste~Olloqui$^{35}$, 
B.~Pie~Valls$^{35}$, 
D.~Piedigrossi$^{37}$, 
B.~Pietrzyk$^{4}$, 
D.~Pinci$^{22}$, 
S.~Playfer$^{46}$, 
M.~Plo~Casasus$^{36}$, 
M.~Poli-Lener$^{18}$, 
G.~Polok$^{25}$, 
A.~Poluektov$^{44,33}$, 
E.~Polycarpo$^{2}$, 
D.~Popov$^{10}$, 
B.~Popovici$^{28}$, 
S.~Poss$^{6}$, 
C.~Potterat$^{38}$, 
A.~Powell$^{51}$, 
S.~Pozzi$^{16,d}$, 
T.~du~Pree$^{23}$, 
V.~Pugatch$^{41}$, 
A.~Puig~Navarro$^{35}$, 
W.~Qian$^{3,7}$, 
J.H.~Rademacker$^{42}$, 
B.~Rakotomiaramanana$^{38}$, 
I.~Raniuk$^{40}$, 
G.~Raven$^{24}$, 
S.~Redford$^{51}$, 
W.~Reece$^{49}$, 
A.C.~dos~Reis$^{1}$, 
S.~Ricciardi$^{45}$, 
J.~Riera$^{35,m}$, 
K.~Rinnert$^{48}$, 
D.A.~Roa~Romero$^{5}$, 
P.~Robbe$^{7,37}$, 
E.~Rodrigues$^{47}$, 
F.~Rodrigues$^{2}$, 
C.~Rodriguez~Cobo$^{36}$, 
P.~Rodriguez~Perez$^{36}$, 
G.J.~Rogers$^{43}$, 
V.~Romanovsky$^{34}$, 
E.~Rondan~Sanabria$^{1}$, 
M.~Rosello$^{35,m}$, 
G.~Rospabe$^{4}$, 
J.~Rouvinet$^{38}$, 
L.~Roy$^{37}$, 
T.~Ruf$^{37}$, 
H.~Ruiz$^{35}$, 
C.~Rummel$^{11}$, 
V.~Rusinov$^{30}$, 
G.~Sabatino$^{21,j}$, 
J.J.~Saborido~Silva$^{36}$, 
N.~Sagidova$^{29}$, 
P.~Sail$^{47}$, 
B.~Saitta$^{15,c}$, 
T.~Sakhelashvili$^{39}$, 
C.~Salzmann$^{39}$, 
A.~Sambade~Varela$^{37}$, 
M.~Sannino$^{19,h}$, 
R.~Santacesaria$^{22}$, 
R.~Santinelli$^{37}$, 
E.~Santovetti$^{21,j}$, 
M.~Sapunov$^{6}$, 
A.~Sarti$^{18}$, 
C.~Satriano$^{22,l}$, 
A.~Satta$^{21}$, 
T.~Savidge$^{49}$, 
M.~Savrie$^{16,d}$, 
D.~Savrina$^{30}$, 
P.~Schaack$^{49}$, 
M.~Schiller$^{11}$, 
S.~Schleich$^{9}$, 
M.~Schmelling$^{10}$, 
B.~Schmidt$^{37}$, 
O.~Schneider$^{38}$, 
T.~Schneider$^{37}$, 
A.~Schopper$^{37}$, 
M.-H.~Schune$^{7}$, 
R.~Schwemmer$^{37}$, 
A.~Sciubba$^{18,k}$, 
M.~Seco$^{36}$, 
A.~Semennikov$^{30}$, 
K.~Senderowska$^{26}$, 
N.~Serra$^{23}$, 
J.~Serrano$^{6}$, 
B.~Shao$^{3}$, 
M.~Shapkin$^{34}$, 
I.~Shapoval$^{40,37}$, 
P.~Shatalov$^{30}$, 
Y.~Shcheglov$^{29}$, 
T.~Shears$^{48}$, 
L.~Shekhtman$^{33}$, 
V.~Shevchenko$^{30}$, 
A.~Shires$^{49}$, 
S.~Sigurdsson$^{43}$, 
E.~Simioni$^{24}$, 
H.P.~Skottowe$^{43}$, 
T.~Skwarnicki$^{52}$, 
N.~Smale$^{10,51}$, 
A.~Smith$^{37}$, 
A.C.~Smith$^{37}$, 
N.A.~Smith$^{48}$, 
K.~Sobczak$^{5}$, 
F.J.P.~Soler$^{47}$, 
A.~Solomin$^{42}$, 
P.~Somogy$^{37}$, 
F.~Soomro$^{49}$, 
B.~Souza~De~Paula$^{2}$, 
B.~Spaan$^{9}$, 
A.~Sparkes$^{46}$, 
E.~Spiridenkov$^{29}$, 
P.~Spradlin$^{51}$, 
A.~Srednicki$^{27}$, 
F.~Stagni$^{37}$, 
S.~Stahl$^{11}$, 
S.~Steiner$^{39}$, 
O.~Steinkamp$^{39}$, 
O.~Stenyakin$^{34}$, 
S.~Stoica$^{28}$, 
S.~Stone$^{52}$, 
B.~Storaci$^{23}$, 
U.~Straumann$^{39}$, 
N.~Styles$^{46}$, 
M.~Szczekowski$^{27}$, 
P.~Szczypka$^{38}$, 
T.~Szumlak$^{47,26}$, 
S.~T'Jampens$^{4}$, 
E.~Tarkovskiy$^{30}$, 
E.~Teodorescu$^{28}$, 
H.~Terrier$^{23}$, 
F.~Teubert$^{37}$, 
C.~Thomas$^{51,45}$, 
E.~Thomas$^{37}$, 
J.~van~Tilburg$^{39}$, 
V.~Tisserand$^{4}$, 
M.~Tobin$^{39}$, 
S.~Topp-Joergensen$^{51}$, 
M.T.~Tran$^{38}$, 
S.~Traynor$^{12}$, 
U.~Trunk$^{10}$, 
A.~Tsaregorodtsev$^{6}$, 
N.~Tuning$^{23}$, 
A.~Ukleja$^{27}$, 
O.~Ullaland$^{37}$, 
U.~Uwer$^{11}$, 
V.~Vagnoni$^{14}$, 
G.~Valenti$^{14}$, 
A.~Van~Lysebetten$^{23}$, 
R.~Vazquez~Gomez$^{35}$, 
P.~Vazquez~Regueiro$^{36}$, 
S.~Vecchi$^{16}$, 
J.J.~Velthuis$^{42}$, 
M.~Veltri$^{17,f}$, 
K.~Vervink$^{37}$, 
B.~Viaud$^{7}$, 
I.~Videau$^{7}$, 
D.~Vieira$^{2}$, 
X.~Vilasis-Cardona$^{35,m}$, 
J.~Visniakov$^{36}$, 
A.~Vollhardt$^{39}$, 
D.~Volyanskyy$^{39}$, 
D.~Voong$^{42}$, 
A.~Vorobyev$^{29}$, 
An.~Vorobyev$^{29}$, 
H.~Voss$^{10}$, 
K.~Wacker$^{9}$, 
S.~Wandernoth$^{11}$, 
J.~Wang$^{52}$, 
D.R.~Ward$^{43}$, 
A.D.~Webber$^{50}$, 
D.~Websdale$^{49}$, 
M.~Whitehead$^{44}$, 
D.~Wiedner$^{11}$, 
L.~Wiggers$^{23}$, 
G.~Wilkinson$^{51}$, 
M.P.~Williams$^{44}$, 
M.~Williams$^{49}$, 
F.F.~Wilson$^{45}$, 
J.~Wishahi$^{9}$, 
M.~Witek$^{25}$, 
W.~Witzeling$^{37}$, 
M.L.~Woodward$^{45}$, 
S.A.~Wotton$^{43}$, 
K.~Wyllie$^{37}$, 
Y.~Xie$^{46}$, 
F.~Xing$^{51}$, 
Z.~Yang$^{3}$, 
G.~Ybeles~Smit$^{23}$, 
R.~Young$^{46}$, 
O.~Yushchenko$^{34}$, 
M.~Zeng$^{3}$, 
L.~Zhang$^{52}$, 
Y.~Zhang$^{3}$, 
A.~Zhelezov$^{11}$ and 
E.~Zverev$^{31}$.\bigskip\newline{\it
\footnotesize
$\dagger$ deceased\\[1ex]
$ ^{1}$Centro Brasileiro de Pesquisas F\'isicas (CBPF), Rio de Janeiro, Brazil\\
$ ^{2}$Universidade Federal do Rio de Janeiro (UFRJ), Rio de Janeiro, Brazil\\
$ ^{3}$Center for High Energy Physics, Tsinghua University, Beijing, China\\
$ ^{4}$LAPP, Universit\'e de Savoie, CNRS/IN2P3, Annecy-Le-Vieux, France\\
$ ^{5}$Clermont Universit\'e, Universit\'e Blaise Pascal, CNRS/IN2P3, LPC, Clermont-Ferrand, France\\
$ ^{6}$CPPM, Aix-Marseille Universit\'e, CNRS/IN2P3, Marseille, France\\
$ ^{7}$LAL, Universit\'e Paris-Sud, CNRS/IN2P3, Orsay, France\\
$ ^{8}$LPNHE, Universit\'e Pierre et Marie Curie, Universit\'e Paris Diderot, CNRS/IN2P3, Paris, France\\
$ ^{9}$Fakult\"at Physik, Technische Universit\"at Dortmund, Dortmund, Germany\\
$ ^{10}$Max-Planck-Institut f\"ur Kernphysik (MPIK), Heidelberg, Germany\\
$ ^{11}$Physikalisches Institut, Ruprecht-Karls-Universit\"at Heidelberg, Heidelberg, Germany\\
$ ^{12}$School of Physics, University College Dublin, Dublin, Ireland\\
$ ^{13}$Sezione INFN di Bari, Bari, Italy\\
$ ^{14}$Sezione INFN di Bologna, Bologna, Italy\\
$ ^{15}$Sezione INFN di Cagliari, Cagliari, Italy\\
$ ^{16}$Sezione INFN di Ferrara, Ferrara, Italy\\
$ ^{17}$Sezione INFN di Firenze, Firenze, Italy\\
$ ^{18}$Laboratori Nazionali dell'INFN di Frascati, Frascati, Italy\\
$ ^{19}$Sezione INFN di Genova, Genova, Italy\\
$ ^{20}$Sezione INFN di Milano Bicocca, Milano, Italy\\
$ ^{21}$Sezione INFN di Roma Tor Vergata, Roma, Italy\\
$ ^{22}$Sezione INFN di Roma Sapienza, Roma, Italy\\
$ ^{23}$Nikhef National Institute for Subatomic Physics, Amsterdam, Netherlands\\
$ ^{24}$Nikhef National Institute for Subatomic Physics and Vrije Universiteit, Amsterdam, Netherlands\\
$ ^{25}$Henryk Niewodniczanski Institute of Nuclear Physics, Polish Academy of Sciences, Cracow, Poland\\
$ ^{26}$Faculty of Physics \& Applied Computer Science, Cracow, Poland\\
$ ^{27}$Soltan Institute for Nuclear Studies, Warsaw, Poland\\
$ ^{28}$Horia Hulubei National Institute of Physics and Nuclear Engineering, Bucharest-Magurele, Romania\\
$ ^{29}$Petersburg Nuclear Physics Institute (PNPI), Gatchina, Russia\\
$ ^{30}$Institute of Theoretical and Experimental Physics (ITEP), Moscow, Russia\\
$ ^{31}$Institute of Nuclear Physics, Moscow State University (SINP MSU), Moscow, Russia\\
$ ^{32}$Institute for Nuclear Research of the Russian Academy of Sciences (INR RAN), Moscow, Russia\\
$ ^{33}$Budker Institute of Nuclear Physics (BINP), Novosibirsk, Russia\\
$ ^{34}$Institute for High Energy Physics (IHEP), Protvino, Russia\\
$ ^{35}$Universitat de Barcelona, Barcelona, Spain\\
$ ^{36}$Universidad de Santiago de Compostela, Santiago de Compostela, Spain\\
$ ^{37}$European Organization for Nuclear Research (CERN), Geneva, Switzerland\\
$ ^{38}$Ecole Polytechnique F\'ed\'erale de Lausanne (EPFL), Lausanne, Switzerland\\
$ ^{39}$Physik Institut, Universit\"at Z\"urich, Z\"urich, Switzerland\\
$ ^{40}$NSC Kharkiv Institute of Physics and Technology (NSC KIPT), Kharkiv, Ukraine\\
$ ^{41}$Institute for Nuclear Research of the National Academy of Sciences (KINR), Kyiv, Ukraine\\
$ ^{42}$H.H. Wills Physics Laboratory, University of Bristol, Bristol, United Kingdom\\
$ ^{43}$Cavendish Laboratory, University of Cambridge, Cambridge, United Kingdom\\
$ ^{44}$Department of Physics, University of Warwick, Coventry, United Kingdom\\
$ ^{45}$STFC Rutherford Appleton Laboratory, Didcot, United Kingdom\\
$ ^{46}$School of Physics and Astronomy, University of Edinburgh, Edinburgh, United Kingdom\\
$ ^{47}$School of Physics and Astronomy, University of Glasgow, Glasgow, United Kingdom\\
$ ^{48}$Oliver Lodge Laboratory, University of Liverpool, Liverpool, United Kingdom\\
$ ^{49}$Imperial College London, London, United Kingdom\\
$ ^{50}$School of Physics and Astronomy, University of Manchester, Manchester, United Kingdom\\
$ ^{51}$Department of Physics, University of Oxford, Oxford, United Kingdom\\
$ ^{52}$Syracuse University, Syracuse, NY, United States\\
$ ^{53}$CC-IN2P3, CNRS/IN2P3, Lyon-Villeurbanne, France, associated member\\
$ ^{54}$Pontif\'icia Universidade Cat\'olica do Rio de Janeiro (PUC-Rio), Rio de
Janeiro, Brazil, associated to $^2 $\\[1ex]
$ ^{a}$Universit\`a di Bari, Bari, Italy\\
$ ^{b}$Universit\`a di Bologna, Bologna, Italy\\
$ ^{c}$Universit\`a di Cagliari, Cagliari, Italy\\
$ ^{d}$Universit\`a di Ferrara, Ferrara, Italy\\
$ ^{e}$Universit\`a di Firenze, Firenze, Italy\\
$ ^{f}$Universit\`a di Urbino, Urbino, Italy\\
$ ^{g}$Universit\`a di Modena e Reggio Emilia, Modena, Italy\\
$ ^{h}$Universit\`a di Genova, Genova, Italy\\
$ ^{i}$Universit\`a di Milano Bicocca, Milano, Italy\\
$ ^{j}$Universit\`a di Roma Tor Vergata, Roma, Italy\\
$ ^{k}$Universit\`a di Roma La Sapienza, Roma, Italy\\
$ ^{l}$Universit\`a della Basilicata, Potenza, Italy\\
$ ^{m}$LIFAELS, La Salle, Universitat Ramon Llull, Barcelona, Spain\\
$ ^{n}$Instituci\'o Catalana de Recerca i Estudis Avan\c cats (ICREA), Barcelona, Spain\\
}
\end{flushleft}

\mbox{~}\vfill
\noindent
{Dedicated to the memory of Werner Ruckstuhl, Peter Schlein and Tom Ypsilantis,\\
who each played a fundamental role in the design of the experiment.}
\vfill\mbox{~}

\cleardoublepage
\setcounter{page}{1}
\pagenumbering{arabic}

\section{Introduction}

Strangeness production studies provide sensitive tests of soft hadronic
interactions, as the mass of the strange quark is of the order of
$\Lambda_{\rm QCD}$. Strange-hadron production is suppressed, as a consequence,
but still occurs in the non-perturbative regime.
The hadronic production of \Kshort\ mesons has been 
studied by several experiments at a range of different
centre-of-mass energies, both in $pp$ and $p\bar{p}$ collisions
(see for example~\cite{ISR,UA5KS2,UA5KS1,CDFKSabs,UA1KS,CDFKS,STARKS}).
The most recent measurements of \Kshort\ production at the Tevatron have shown deviations 
with respect to the expectations of hadronization models~\cite{CDFKS}.  
Strangeness production is also a topic of great interest in heavy ion physics, and
measurements of this process in $pp$ and $p\bar{p}$ collisions serve as reference point~\cite{STARKS}.

In this paper measurements of prompt \Kshort\ production are presented using
data collected with the LHCb detector in $pp$ collisions at $\sqrt{s}=0.9$~TeV,
during the 2009 pilot run of the Large Hadron Collider (LHC).  A \Kshort\ is
defined to be prompt if it is directly produced in the $pp$ collision, 
or if it appears in the decay chain of a
non-weakly-decaying resonance (such as $K^*$) directly produced in the $pp$ collision.
The measurements are made in the rapidity interval $2.5 < y < 4.0$
and down to below 0.2~GeV$/c$ transverse momentum with respect to the beam line. This 
is a region not explored at this energy by any previous experiment,
and is complementary to the coverage of other LHC experiments.
The determination of the prompt \Kshort\
production cross-section is normalized using
an absolute measurement of the luminosity that relies on 
knowledge of the beam profiles.

The paper is organized as follows.
Section~\ref{sec:detector} gives a brief description of the LHCb detector and the 
configuration used to record data in December 2009 during the LHC pilot run. 
Section~\ref{sec:strategy} gives an overview of the analysis
strategy, the details of which are presented in the three 
following sections. Section~\ref{sec:luminosity} is 
dedicated to an explanation of the luminosity
measurement, Section~\ref{sec:selection} presents the \Kshort\ candidate
selection and Section~\ref{sec:efficiency} the determination of the
\Kshort\ trigger and reconstruction efficiencies. The final results are discussed  
in Section~\ref{sec:results} 
and compared with 
model expectations, before concluding in Section~\ref{sec:conclusion}.

\section{LHCb detector and 2009 data sample}
\label{sec:detector}

The LHCb detector is a single-arm magnetic dipole spectrometer with a polar angular coverage
with respect to the beam line of approximately 15 to 300~mrad in the horizontal
bending plane, and 15 to 250~mrad in the vertical non-bending plane.
The detector is described in detail elsewhere~\cite{LHCb}.
All subdetectors were fully operational and in a stable condition for the
data that are analysed.
For the measurements presented in
this paper the tracking detectors and trigger strategy are of particular importance.

A right-handed coordinate system is defined with its origin at the 
nominal $pp$ interaction point, the $z$ axis along the beam
line and pointing towards the magnet, and the $y$ axis pointing upwards. 
Beam-1 (beam-2) travels in the direction of positive (negative) $z$.

The LHCb tracking system consists of the Vertex Locator (VELO) 
surrounding the $pp$ interaction region, a 
tracking station (TT) upstream of the dipole magnet, and three tracking stations
(T1--T3) downstream of the magnet.  Particles traversing from the interaction region
to the downstream tracking stations experience a bending-field integral of 3.7~Tm on average.

The VELO consists of silicon microstrip modules,
providing a measure of the radial and azimuthal coordinates, 
$r$ and $\phi$, distributed in 23 stations arranged
along the beam direction.
The first two stations at the most upstream $z$ positions
are instrumented to provide
information on the number of visible interactions in the detector at the
first level of the trigger (`pile-up detector').
The VELO is constructed in two halves (left and right),
movable in the $x$ and $y$ directions so that it can be centred on the beam. 
During stable beam conditions the two halves 
are located at their nominal closed position, with active silicon at 
8~mm from the beams,
providing full azimuthal coverage.
During injection and beam adjustments
the two halves are moved apart horizontally 
to a retracted position away from the beams. 

The TT station also uses silicon microstrip technology. The downstream tracking stations T1--T3
have silicon microstrips in the region close to the beam pipe  (Inner Tracker, IT), whereas
straw tubes are employed in the outer region (Outer Tracker, OT).

During the 2009 run, low intensity beams collided in LHCb at the LHC injection
energy, corresponding to a total energy of 0.9~TeV. Due to the dipole magnetic
field the beams have a crossing angle that results in the $pp$
centre-of-mass frame moving with velocity $0.0021c$ in the $-x$ direction.
Both the beam sizes and crossing angle 
were larger than those designed for high-energy collisions.
In order not to risk the safety of the VELO,
the 2009 data were recorded with the 
two VELO halves positioned 15~mm
away from their nominal data-taking position (VELO partially open), 
resulting in a reduced azimuthal coverage.
For this run, the magnetic dipole field was pointing downwards. 

The bulk of the data presented here were collected in a series of LHC fills with
the following two sets of beam conditions. The first configuration contained four
bunches per beam, spaced by more than 8 $\mu$s, with two colliding and two
non-colliding bunches, and a total peak beam intensity of about
$1.8\times 10^{10}$ protons per bunch. The second configuration contained 16
bunches per beam, spaced by more than 2 $\mu$s, with eight colliding and eight
non-colliding bunches, and a total peak beam intensity of about
$1.3\times 10^{10}$ protons per bunch.  The nominal LHC injection optical
function at the interaction point was used ($\beta^* = 10$~m).

A trigger strategy was deployed to provide high efficiency
for $pp$ inelastic interactions and for beam collisions with
the residual gas in the vacuum chamber.
The latter class of events is a necessary ingredient
for the luminosity analysis. Events were collected for three
bunch-crossing types:
two colliding bunches ($bb$), beam-1 bunch with no beam-2 bunch ($b1$),
and beam-2 bunch with no beam-1 bunch ($b2$). 
The first two categories of crossings,
which produce particles in the forward ($+z$) direction, were triggered using
calorimeter information: a 2$\times$2 cluster with more than 240~MeV
of transverse energy in the Hadron Calorimeter (HCAL) 
and at least three hits in the 6016 cells
of the Scintillator Pad Detector (SPD) at the entrance to the
calorimeter were required. Events containing a track in
the muon system with transverse momentum above 480~MeV$/c$
were also triggered. Crossings of the type $b2$, which produce particles 
in the backward direction only, were 
triggered by demanding a hit multiplicity of more than seven
in the pile-up detector.

The visible collision rate for a single bunch pair was about 10 Hz and the
acquired $b1$ ($b2$) rate for a single
bunch was approximately 0.015~Hz (0.002~Hz), in agreement with the 
measured residual pressure and VELO acceptance.
A sample of $424\,193$ events triggered in $bb$ crossings is used in the \Kshort\ analysis.

\section{Analysis strategy}
\label{sec:strategy}

All \Kshort\ candidates are reconstructed in the $\pi^+\pi^-$ decay mode, 
using only events triggered by the calorimeter. 
Contributions from secondary interactions in the 
detector material or from the decay of long-lived
particles are suppressed by 
requiring the \Kshort\ candidates to point back to the $pp$-collision point. 
No attempt is made to separate the contributions
from \Kshort\ mesons produced in diffractive and non-diffractive processes.

Due to the long \Kshort\ lifetime and partially open VELO position, 
only a small fraction of the \Kshort\ daughter tracks traversing the 
spectrometer leave a signal in the VELO. 
Therefore, two paths are followed
for the \Kshort\ reconstruction and selection:
\begin{itemize}
\item[a)] Downstream-track selection: \newline 
Tracks reconstructed only with hits in the TT and T1--T3 stations
(called downstream tracks) are combined,
without using the VELO. 
The origin of the \Kshort\ is taken as
the point on the $z$ axis that is closest
to the reconstructed flight vector
of the \Kshort\ candidate.
This point is taken as an estimate of the primary vertex (PV),
and is referred to as the `pseudo-PV'.
\item[b)] Long-track selection: \newline
\Kshort\ candidates are formed with tracks
leaving hits in the VELO and in the T stations (called long tracks). 
If available,  measurements in the TT are added to the tracks.
The PV is reconstructed from tracks seen in the detector, 
using VELO information whenever available. 
\end{itemize}

The analysis is performed in bins of \Kshort\ phase space. 
The kinematic variables used are
the \Kshort\ transverse momentum $p_{\rm T} = \sqrt{p_x^2+p_y^2}$
and the rapidity 
$y = \frac{1}{2}\ln((E+p_z)/(E-p_z))$,
where $(E,\vec{p}\,)$ is the \Kshort\ four-momentum in the $pp$ centre-of-mass system.
For a given bin $i$ in $p_{\rm T}$ and $y$,
the prompt \Kshort\ production cross-section is calculated as 
\begin{equation}
\sigma_i = \frac{N_i^{\rm obs}}{\epsilon_i^{\rm trig/sel} ~ \epsilon_i^{\rm sel} ~ L_{\rm int}}\,,
\label{eq:sigma_i}
\end{equation}
where $N_i^{\rm obs}$ is the number of observed $\Kshort \to \pi^+\pi^-$ signal 
decays with reconstructed $p_{\rm T}$ and $y$ in bin $i$,
$\epsilon_i^{\rm sel}$ the reconstruction and selection efficiency, 
$\epsilon_i^{\rm trig/sel}$ the trigger efficiency on selected events, 
and $L_{\rm int}$ the integrated luminosity.
The number of signal events $N_i^{\rm obs}$ is obtained
from the mass distributions of the \Kshort\ candidates.

The reconstruction and selection efficiency
is estimated from a fully-simulated Monte Carlo (MC)
sample of single $pp$ collisions as
\begin{equation}
\epsilon_i^{\rm sel} = \frac{N_i^{\rm sel}}{N_i^{\rm prompt}}\,,
\label{eq:epsilon_i_sel}
\end{equation}
where $N_i^{\rm sel}$ is the number of $\Kshort \to \pi^+\pi^-$ signal decays selected
in the untriggered MC sample with reconstructed $p_{\rm T}$ and $y$ in bin $i$
(extracted using the same procedure as in the data),
and where $N_i^{\rm prompt}$ is the number of generated prompt \Kshort\ mesons 
with generated $p_{\rm T}$ and $y$ in bin $i$.
This efficiency includes the geometrical acceptance, 
as well as the reconstruction and selection efficiencies. 
It also incorporates all corrections related to the following effects: 
secondary interactions of \Kshort\ in the material, 
$\Kshort \to \pi^+\pi^-$ branching fraction, decay in flight and secondary 
interaction of the decay products, non-prompt \Kshort\ production and finite 
resolution of the $p_{\rm T}$ and $y$ observables. 
 
The trigger efficiency is estimated using the same MC events. However, 
since the efficiency depends on the global event properties, the MC events 
are weighted to reproduce the observed track multiplicity in the selected 
signal events. Then 
\begin{equation}
\epsilon_i^{\rm trig/sel} = \frac{Y_i^{\rm trig/sel}}{Y_i^{\rm sel}}
\label{eq:epsilon_i_trig}
\end{equation}
is computed, where $Y_i^{\rm trig/sel}$ and $Y_i^{\rm sel}$
are the weighted MC signal yields extracted
after and before the trigger cuts are applied. 

The integrated luminosity $L_{\rm int}$ is determined using a
novel `beam imaging' method~\cite{Massi}, taking advantage of proton
collisions with the 
residual gas in the interaction region and of the excellent
vertexing capability of the VELO. The beam profiles 
and positions are reconstructed using tracks
produced in beam-gas and beam-beam collisions.
Combining this information with bunch current measurements 
from the LHC machine yields a direct 
measurement of the integrated luminosity. 

\section{Luminosity determination}
\label{sec:luminosity}

In the relativistic approximation, the average instantaneous luminosity 
produced by one pair of colliding bunches can be expressed as~\cite{Moller}

\begin{equation}\label{eq:luminosity}
L =  2 \, c\,n_1\, n_2 \, f \, \cos^2{\theta} \int{\rho_1(x,y,z,t)\rho_2(x,y,z,t) \ {\rm d}x\,{\rm d}y\,{\rm d}z\,{\rm d}t} \ ,
\end{equation}
where $n_i$ are the number of protons in bunch $i$ ($i=1,2$),
$f=11.245~{\rm kHz}$ is the LHC revolution frequency, 
$\theta$ is the half crossing angle of the beams,
and $\rho_i(x,y,z,t)$ is the 
density of bunch $i$ normalized as
$\int \rho_i(x,y,z,t)\, {\rm d}x\, {\rm d}y\, {\rm d}z=1$
at all times $t$.
The overlap integral in Eq.~(\ref{eq:luminosity}) 
is taken over the duration of one bunch crossing. 
Tracks measured in the VELO allow vertices from beam-gas and beam-beam 
collisions to be reconstructed for each pair of bunches.
From the distributions of these
vertices, and assuming the gas density to be uniform
in any plane transverse to the beams,
the positions, angles and sizes of the
bunches are measured, and their overlap integral is computed.
The numbers of protons per bunch are determined 
with the LHC machine instrumentation, 
enabling an absolute normalization of the luminosity.
The total luminosity is then obtained as the sum of the 
estimates 
for each pair of colliding bunches in the machine. 

The beam crossing angle is limited to the 
horizontal plane.
No correlation between the 
transverse coordinates 
is observed at the level of precision needed for this analysis, thus the
$x$ and $y$ projections can be factorized.
The bunch shapes are well described by Gaussian distributions
in all three dimensions, characterized in the $x-y$ plane at the 
time of crossing by their width $\sigma_{ij}$ 
and their mean position $\mu_{ij}$ ($j=x, y$), and by 
their average longitudinal width $\sigma_z$, assumed to be equal for both beams. With these 
approximations and for small crossing angle, Eq.~(\ref{eq:luminosity}) can be rewritten as 
\begin{equation}
L=\frac{n_1 n_2\,f}{2\pi\sqrt{1+2(\theta\sigma_z)^2/(\sigma_{1x}^2 + \sigma_{2x}^2)}} ~
 \prod_{j=x,y} \frac{1}{\sqrt{\sigma_{1j}^2 + \sigma_{2j}^2}} 
 \exp{\left(-\frac{1}{2} \frac{(\mu_{1j}-\mu_{2j})^2}{\sigma_{1j}^2+\sigma_{2j}^2}\right)} \,.
\label{eq:obslumi}
\end{equation}

\begin{figure}[t]
\begin{center}
\epsfig{file=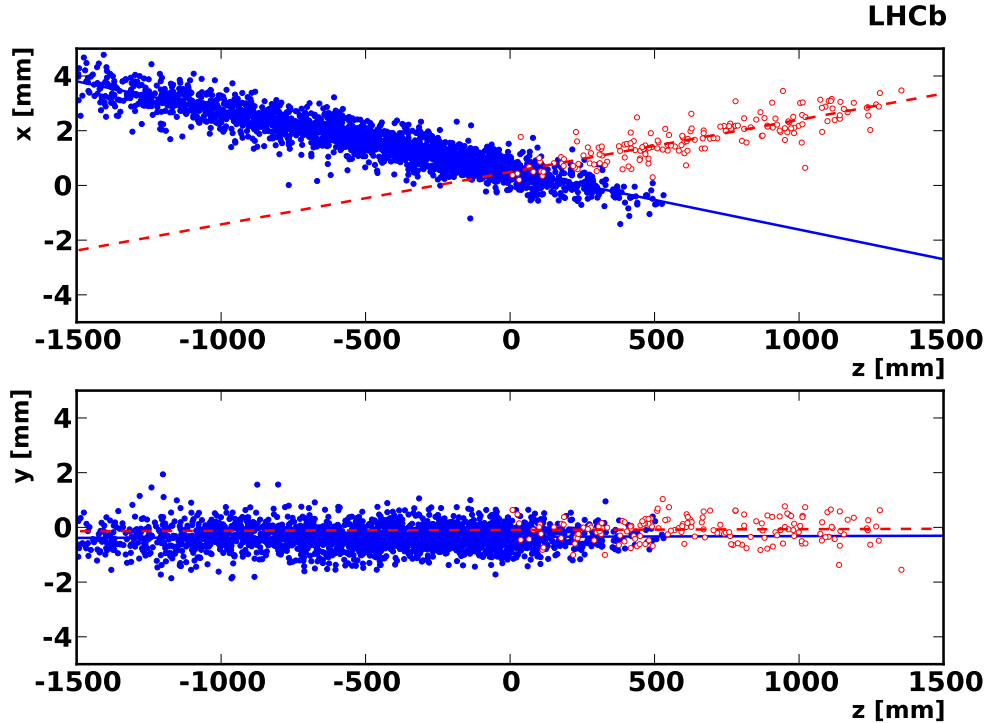,bbllx=25,bblly=180,bburx=594,bbury=580,width=0.8\textwidth}
\caption{\small Distributions in the horizontal (top) and vertical (bottom)
planes of the reconstructed 
vertices in $b1$ (blue filled circles and solid fit line) and $b2$
(red open circles and dashed fit line) crossings in one fill.}
\label{fig:bslopes}
\end{center}
\end{figure}

The observables $\sigma_{ij}$ and $\mu_{ij}$ are extracted 
from the transverse distributions of the beam-gas vertices 
reconstructed in the $bb$ crossings of the colliding bunch pair
with a $z$ coordinate satisfying
$-1000<z<-200$~mm ($200<z<1000$~mm) for $i=1$ ($i=2$). 
These transverse distributions are obtained by projecting 
the reconstructed vertex positions onto a plane perpendicular to 
the corresponding beam direction. 
As illustrated in Fig.~\ref{fig:bslopes},
the beam directions, and hence also the half crossing angle $\theta$, 
are obtained from straight-line fits through the measured 
positions of vertices reconstructed in $b1$ and $b2$ crossings
of other non-colliding bunches. 
The observed half crossing angle of $\theta = 2.1 \pm 0.1$~mrad
in the horizontal plane is 
in agreement with the expected value.

In addition, the distribution of $pp$-collision vertices, produced by the 
colliding bunch pair and identified by requiring $-150<z<150$~mm, 
can be used to measure the parameters of the luminous region. 
Its position $\mu_{j}$ and transverse width $\sigma_{j}$,
 \begin{equation}
\mu_{j} = \frac{\mu_{1j}\sigma_{2j}^2 +
\mu_{2j}\sigma_{1j}^2}{\sigma_{1j}^2 + \sigma_{2j}^2} 
~~~ \mbox{and} ~~~ 
\sigma_{j}^2 = \frac{\sigma_{1j}^2 \sigma_{2j}^2}%
{\sigma_{1j}^2 + \sigma_{2j}^2}  \,,~~(j=x,y)
\label{eq:lumi-gaussian}
\end{equation}
can be used to constrain the bunch observables. 
Owing to the higher statistics of $pp$ collisions
compared to beam-gas interactions,
the constraints of Eq.~(\ref{eq:lumi-gaussian})
provide the most significant input to the overlap integral.

The longitudinal bunch size $\sigma_z$ is extracted
from the longitudinal distribution of the $pp$-collision vertices.
Because $\sigma_z$ is approximately 200 times larger than $\sigma_{ix}$,
the crossing angle reduces the luminosity by a non-negligible
factor equal to the first square root term in Eq.~(\ref{eq:obslumi}).
For the fill used to determine the absolute luminosity, 
this factor is estimated to be $1.087 \pm 0.012$.

\begin{table}[t]
\caption{\small Parameters describing the vertex resolution functions defined
 in Eqs.~(\ref{eq:resolution_function}) and (\ref{eq:resolution_sigma}).
The quoted errors include statistical and systematic uncertainties.
The parameters $f_j$ and $r_j$ were fixed in the fits, and hence have 
no uncertainties.}
\begin{center}
\begin{tabular}{|c|@{\,}c@{\,}|@{\,}c@{\,}|@{\,}c@{\,}|@{\,}c@{\,}|@{\,}c@{\,}|@{\,}c@{\,}|@{\,}c@{\,}|@{\,}c@{\,}|}
\hline
& $f_j$
& $r_j$
& $s_j^{\rm track}~[\rm\mu m]$
& $\delta_j$
& $b_{1j}$
& $m_{1j}~[\rm m^{-1}]$
& $b_{2j}$
& $m_{2j}~[\rm m^{-1}]$ \\ \hline
$x$ & $0.9$ & $0.32$ & $177 \pm 7$ & $5.9 \pm 1.1$ 
    & $1.18 \pm 0.07$   & $-0.86 \pm 0.30$ 
    & $0.83 \pm 0.14$ & $+0.77 \pm 0.24$ \\
$y$ & $0.9$ & $0.36$ & $164 \pm 6$ & $3.7 \pm 1.1$ 
    & $1.24 \pm0.08$ & $-0.57 \pm 0.16$ 
    & $0.85 \pm 0.14$ & $+0.77 \pm 0.24$ \\
\hline
\end{tabular}
\end{center}
\label{tab:lumi_vert_res} 
\end{table}

The vertex resolutions need to be measured
since they are of the same order as the bunch sizes.
This is achieved by comparing,
on an event-by-event basis, the reconstructed vertex positions 
obtained from two independent sets of tracks.
In each event, the sample of available tracks is randomly
split into two sets of equal multiplicity, 
and the event is kept only if exactly one vertex is reconstructed
for each set. In this case the two vertices are assumed to 
originate from the same interaction. 
The vertex resolution for each coordinate is obtained
as the width of the distribution of the difference
in position between the two reconstructed vertices divided by $\sqrt{2}$.
A systematic study of the vertex resolutions in both $x$ and $y$ is then
performed as a function of the number of tracks $N$ contributing to the 
vertex, of the crossing type,
and of the $z$ coordinate of the vertex.
The resolution functions are found to 
be well parametrized by a double Gaussian function
\begin{equation}
R_j(N,z) = f_j ~ G(s_j(N,z))+(1-f_j) ~ G(s_j(N,z)/r_j)
\,,\quad (j =x,  y) \,,
\label{eq:resolution_function}
\end{equation}
where $f_j$ is the fraction of events in the first Gaussian function,
$r_j$ is the ratio of the widths of the two Gaussian functions, 
and $G(s_j(N,z))$ is a Gaussian function centred at zero
with width
\begin{equation}
\begin{array}{lll}
s^{bb}_j(N,z) & =  N^{-0.5+\delta_j/N^2} ~ s_j^{\rm track}  &
\mbox{for beam-beam} \\
s^{i}_j(N,z)  & = (b_{ij} + m_{ij} z) ~ s^{bb}_j(N,z)  & \mbox{for beam-gas ($i=1,2$)}
\end{array}
\,, \quad (j =x,  y) \,.
\label{eq:resolution_sigma}
\end{equation}
The parameters $s_j^{\rm track}$ describe the per-track resolutions,
$\delta_j$ specify the dependence on the number of tracks, while
$b_{ij}$ and $m_{ij}$ model the linear $z$ dependence 
for beam-gas vertices.
The validity of this parametrization has been verified with MC simulation studies.
The systematic uncertainties on the parameters are estimated
from the level of agreement in that check.
The final set of resolution parameters is given
in Table~\ref{tab:lumi_vert_res}.
The resolution is found to be better in $y$ than in $x$, which is 
expected from the partial VELO opening described in
Section~\ref{sec:detector}.

\begin{figure}[t]
\begin{center}
\epsfig{file=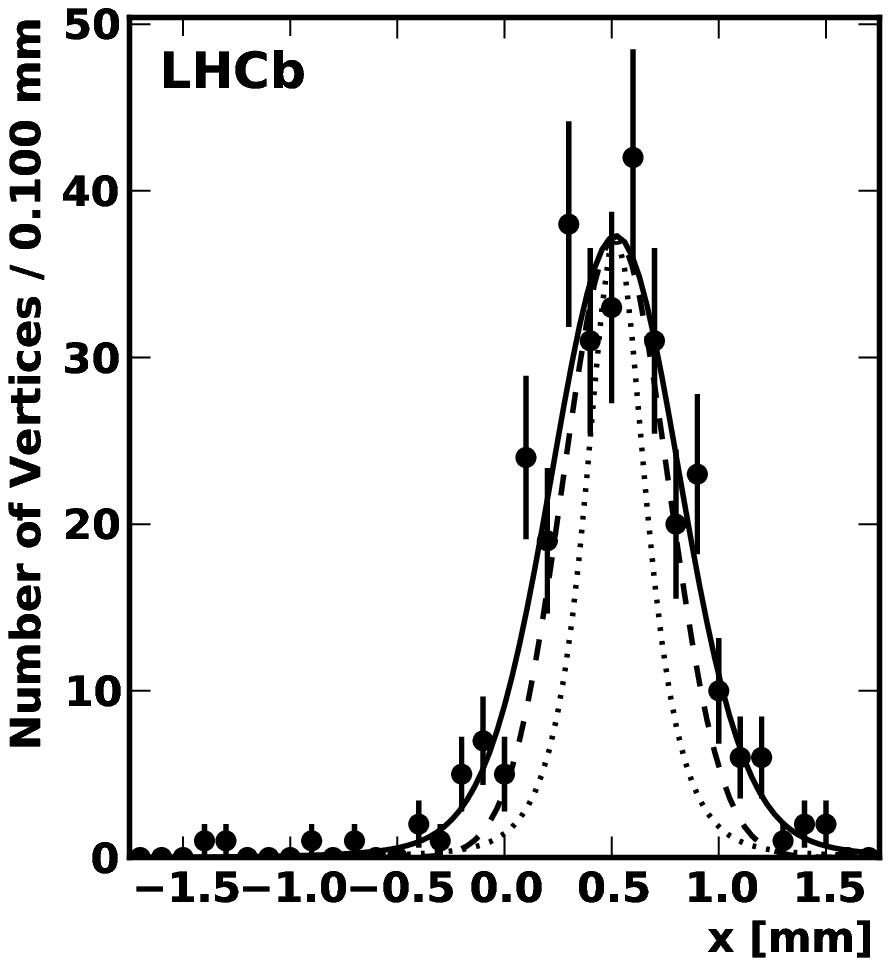,bbllx=180,bblly=260,bburx=438,bbury=535,width=0.32\textwidth} \hfill
\epsfig{file=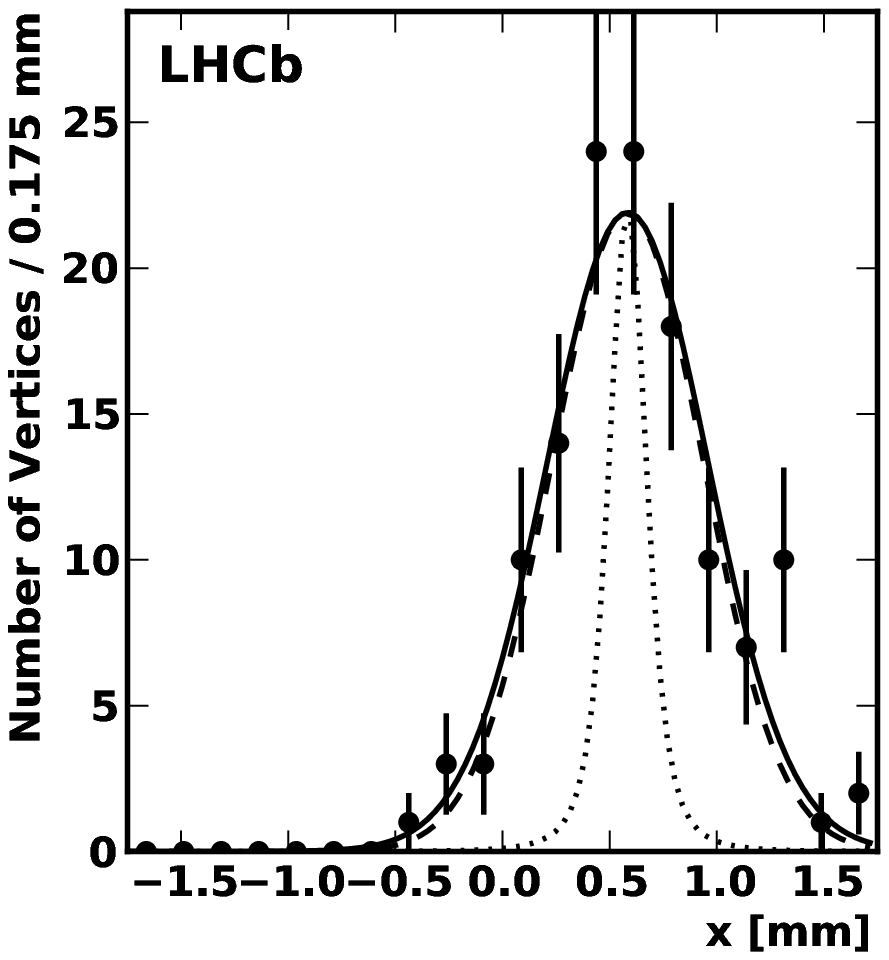,bbllx=180,bblly=260,bburx=438,bbury=535,width=0.32\textwidth} \hfill
\epsfig{file=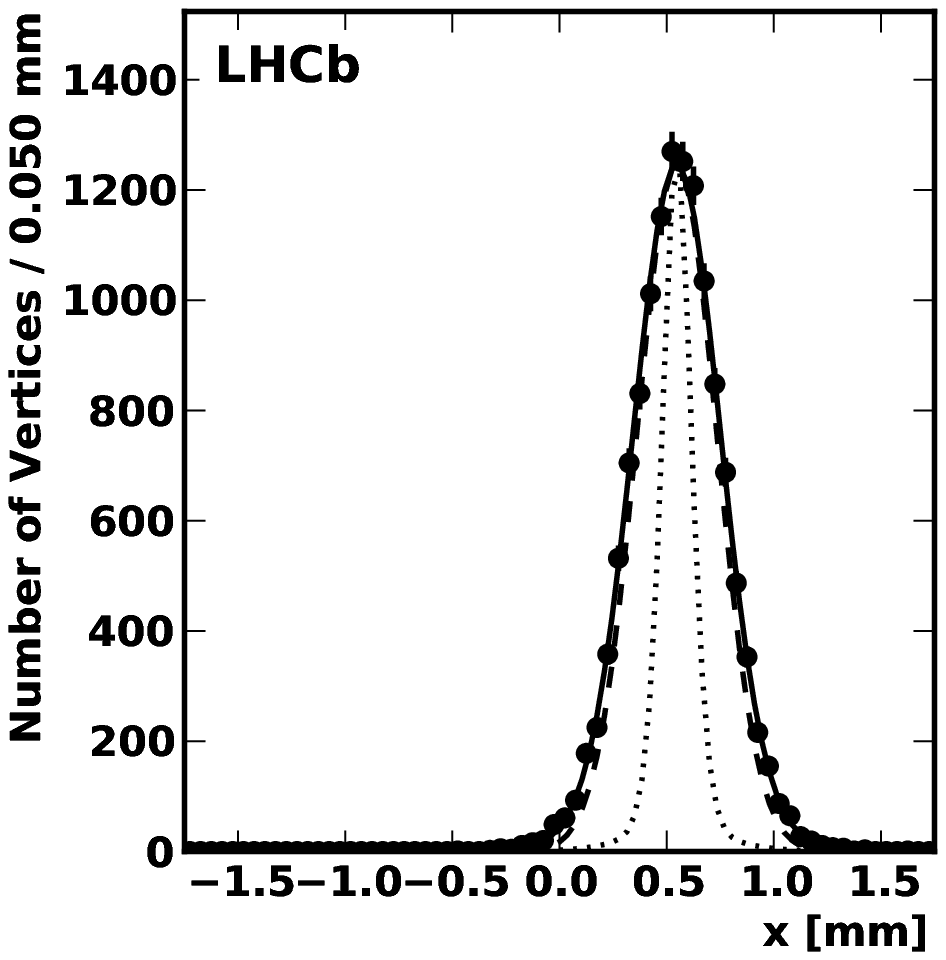,bbllx=180,bblly=260,bburx=438,bbury=535,width=0.32\textwidth}
\end{center}\vspace*{-5mm}
\caption{\small
Measured profiles of beam-1, beam-2 and luminous region (from left to right) 
in the horizontal direction $x$, in $bb$ crossings
of one pair of colliding bunches 
in one fill. The solid curve is a fit to the observed distributions,
the dotted curve represents the vertex resolution, 
and the dashed curve shows the underlying distributions
after deconvolution of the vertex resolution.}
\label{fig:bsizes}
\end{figure}

For both transverse coordinates, each sample of vertices
(defined for every colliding bunch pair in each fill)
is fitted with convolutions of the Gaussian beam shapes
with the resolution function of Eq.~(\ref{eq:resolution_function}). This 
fit is performed with all three types of interactions. With the constraints of Eq.~(\ref{eq:lumi-gaussian}),
this yields directly the position $\mu_{ij}$ and Gaussian
width $\sigma_{ij}$ of the underlying distributions. Some example 
distributions are shown in  Fig.~\ref{fig:bsizes}.
The systematic errors on the results are estimated by varying the 
resolution parameters within their total uncertainties. 

The remaining ingredients needed for the direct
luminosity measurement are the bunch intensities.
The LHC is equipped with two systems
of beam current transformers (BCT)~\cite{BCT}. 
A DC-BCT system provides an ungated measurement of
the total beam current, while a 
fast-BCT system is gated to measure the current induced on a
bunch-by-bunch basis.
The individual bunch intensities are obtained from these fast-BCT 
readings, but constraining
their sum to the DC-BCT measurements.
At the low intensities of the 2009 pilot run, the offset
in the DC-BCT digitization is non-negligible and is corrected
by averaging the readings in the periods without circulating beam just
before and after a fill.

The method described above was used to measure the luminosity in 
four different machine fills.  
Two of those fills were relatively short and the third was  
taken before optimization of the beam alignment. The remaining fill,
taken under optimal conditions and 
representing approximately 25\% of the sample
used for the \Kshort\ production study,
is chosen to determine the absolute normalization of the luminosity for the
data set used for the \Kshort\ analysis. 
The other three fills yield less precise but
consistent results. 
The integrated luminosity for the data set used for the \Kshort\ selection, 
$L_{\rm int}= 6.8 \pm 1.0~\mu{\rm b}^{-1}$, 
is obtained by scaling with the number
of $pp$ interaction vertices measured with the VELO.
The relative uncertainty on this result comprises contributions 
from the measurements of the beam intensities (12\%), widths (5\%), 
relative positions (3\%) and crossing angle (1\%). 
This is the most precise determination of the luminosity for the 2009 LHC
pilot run. The limiting uncertainty on the beam intensity is expected to
improve in the future.

\section{\Kshort\ selection and signal extraction}
\label{sec:selection}

\begin{table}[t]
\caption{\small $\Kshort\to\pi^+\pi^-$ selection requirements. 
}
\begin{center}
\begin{tabular}{|l|l|}
\hline
Variable & Requirement \\ 
\hline 
\multicolumn{2}{|c|}{Downstream-track selection} \\
\hline
Each $\pi$-track momentum             & $> 2~{\rm GeV}/c$ \\
Each $\pi$-track transverse momentum &  $> 0.05~{\rm GeV}/c$ \\
Each track fit $\chi^2$/ndf             & $<25$ \\
Distance of closest approach of each $\pi$-track to the $z$ axis & $> 3$~mm \\
\Kshort\ decay vertex fit $\chi^2$/ndf      & $< 25$ \\
$z$ of \Kshort\ decay vertex        & $< 2200$~mm \\ 
$|z|$ of pseudo-PV & $<150$~mm \\
$\cos \theta_{\rm pointing}$ & $>0.99995$ \\
\Kshort\ proper time ($c\tau$) & $>5$~mm \\
\hline 
\multicolumn{2}{|c|}{Long-track selection} \\
\hline
$|z|$ of associated PV & $< 200$~mm \\
Each track fit $\chi^2$/ndf             & $<25$ \\
\Kshort\ decay vertex $\chi^2$/ndf      & $< 100$ \\
$z(\Kshort)-z({\rm PV})$  & $ > 0$~mm \\
Variable $\nu$ related to impact parameters & $>2$ \\
\hline
\end{tabular}
\end{center}
\label{tab:cuts} 
\end{table}

In the downstream-track selection,
a \Kshort\ candidate is formed from any combination of
two oppositely-charged downstream tracks,
assumed to be pions,
satisfying the requirements listed in the top part of Table~\ref{tab:cuts}.
The pseudo-PV was defined in Section~\ref{sec:strategy}, and 
$\theta_{\rm pointing}$ is the angle between the \Kshort\ momentum vector
and the direction joining the pseudo-PV and the \Kshort\ decay vertex.

In the long-track selection, primary vertices are reconstructed
from at least three tracks. Each 
\Kshort\ candidate formed from long tracks is associated with the PV that minimizes its impact parameter
and the requirements listed in the 
bottom part of Table~\ref{tab:cuts} are applied. 
The variable $\nu$ is 
similar to a Fisher discriminant 
formed with the logarithms of the impact parameters; it is defined as 
$\nu= \ln{[(I_+ ~ I_-)/(I_0 ~ I_1)]}$.
Here $I_+$, $I_-$ and $I_0$ are the
impact parameters of each of the two tracks
and of the \Kshort\ candidate with respect to their closest PV,
respectively, and the value of $I_1$ is fixed to 1~mm.

\begin{figure}[t]
\begin{center}
\epsfig{file=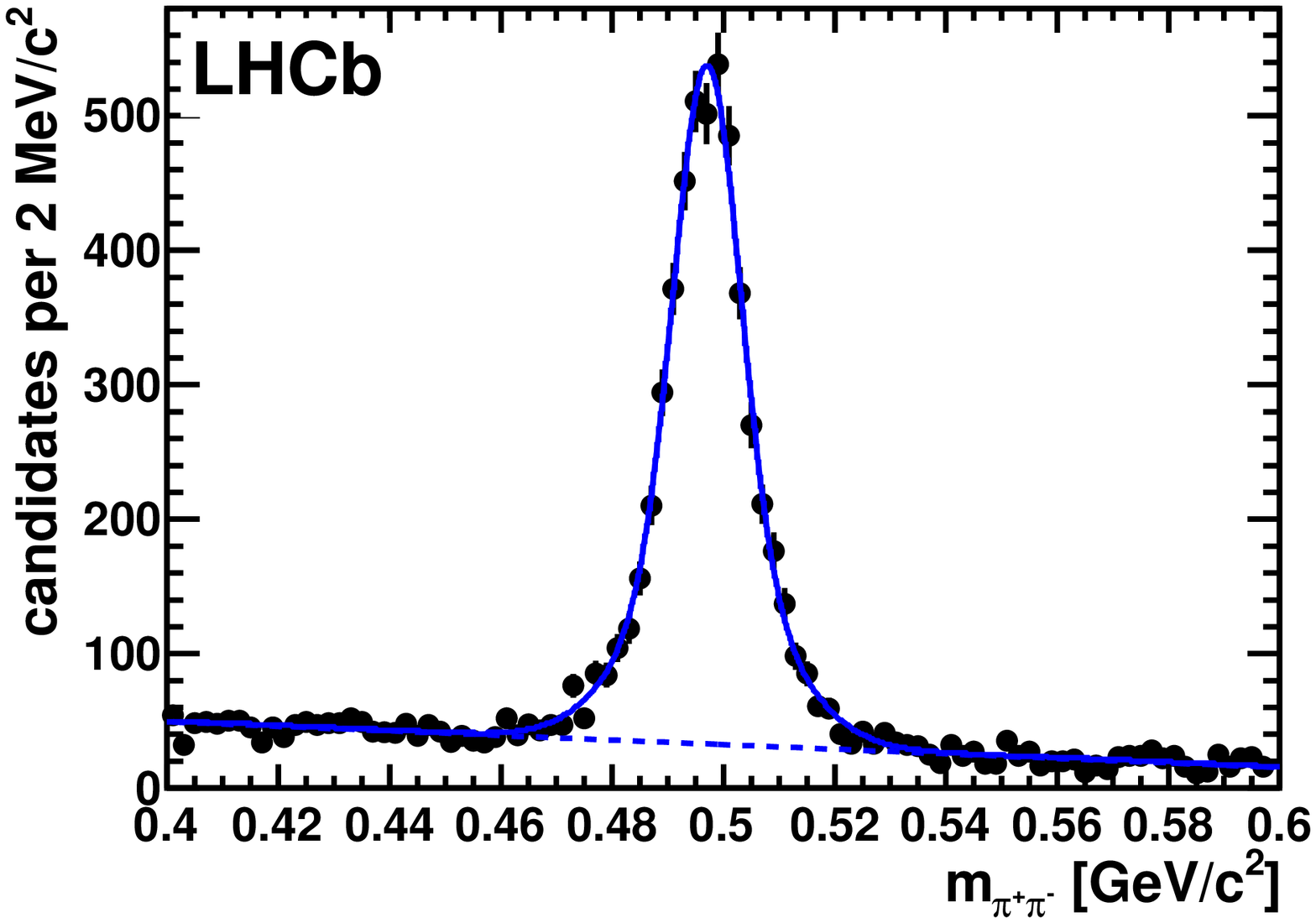,bbllx=10,bblly=10,bburx=553,bbury=390,width=0.49\textwidth}
\epsfig{file=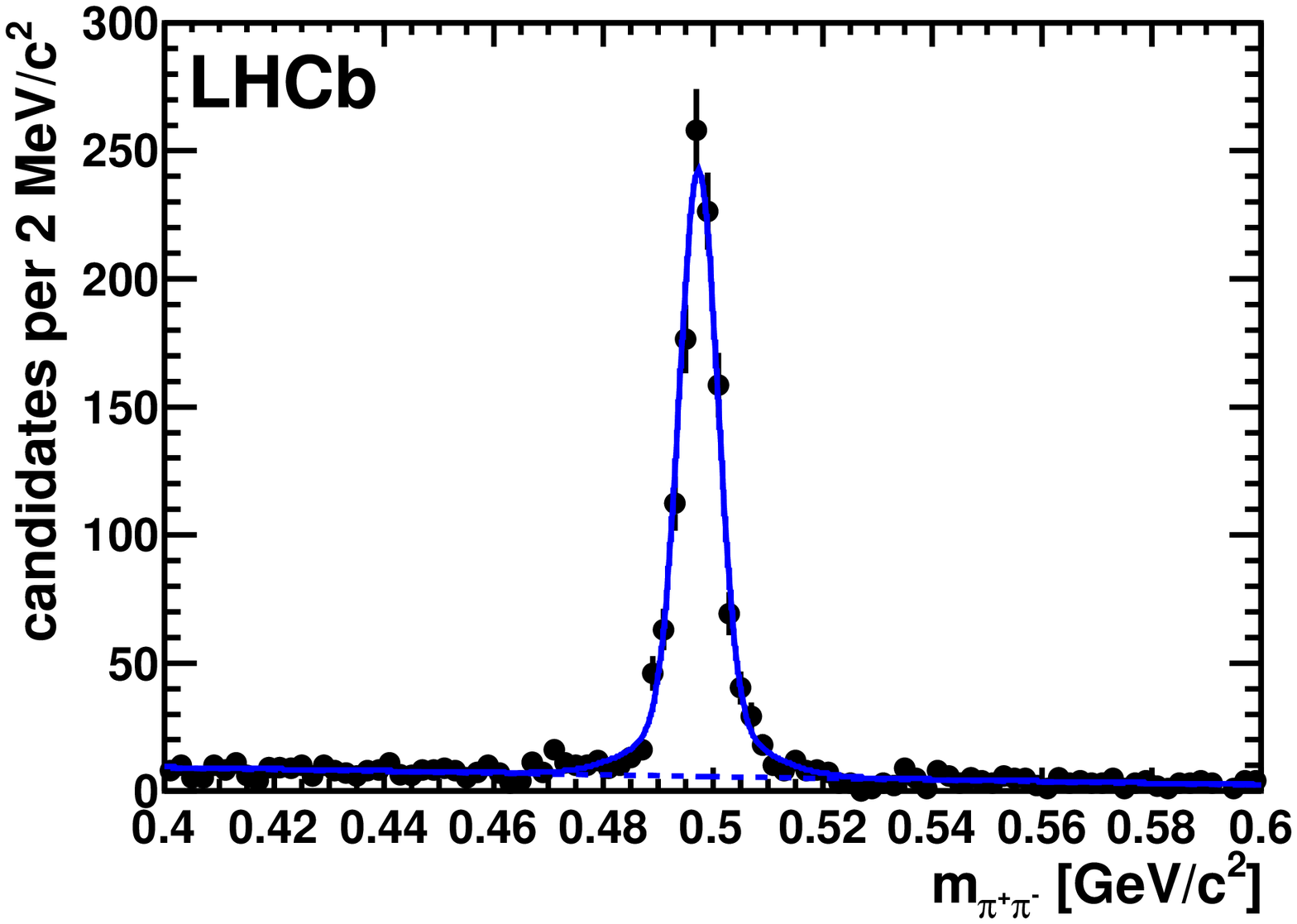,bbllx=10,bblly=10,bburx=553,bbury=390,width=0.49\textwidth}
\caption{\small Mass distributions of all selected \Kshort\ candidates, 
in the downstream-track (left) and long-track (right) selections. 
The points are the beam-gas subtracted data and the curves
are the result of the fits described in the text.}
\label{fig:mass_all}
\end{center}
\end{figure}

Mass distributions are obtained for both $bb$ crossings and $b1$ crossings.
In order to keep only the contribution arising from $pp$ collisions,
the $b1$ mass distribution is subtracted, after proper normalization, 
from the $bb$ mass distribution. 
The normalization factor is $0.908 \pm 0.015$,
averaged over the entire sample used for this analysis. It
is obtained from the ratio of the number of interaction vertices reconstructed
in $bb$ and $b1$ crossings in the region $z< -200$~mm
where no $pp$ collision can take place. This beam-gas subtraction removes
about 1.2\% of the \Kshort\ signal. 

The beam-gas subtracted 
mass distributions are shown in Fig.~\ref{fig:mass_all}
for all selected \Kshort\ candidates. 
A $\chi^2$ fit is made, describing the
background with a linear function and the signal
with the sum of two Gaussian functions of common
mean value, with all parameters left free. It
gives a total \Kshort\ signal yield of
 $4801 \pm 84$   
($1140 \pm 35$), 
a mean mass value of
 $497.12 \pm 0.14~{\rm MeV}/c^2$   
($497.43 \pm 0.14~{\rm MeV}/c^2$), 
and an average resolution of
 $9.2~{\rm MeV}/c^2$  
($5.5~{\rm MeV}/c^2$) 
for the downstream-track (long-track) selection.
Quoted uncertainties are statistical only.
The mass values are close to the known \Kshort\ mass value of \mbox{$497.61 \pm
0.02~{\rm MeV}/c^2$ \cite{PDG},}
reflecting the current status of the mass-scale calibration.
In the long-track selection,
the statistics are lower than in the downstream-track selection,
but the background level is lower
and the mass resolution is significantly better. 

\begin{table}[t] 
\caption{\small 
Number of observed beam-gas subtracted $\Kshort\to\pi^+\pi^-$ signal decays,
as extracted in the downstream- and long-track selections for each bin
of transverse momentum $p_{\rm T}$ and rapidity $y$. 
The first quoted uncertainty is statistical and the second systematic.
The latter is uncorrelated across bins. A dash indicates that the statistics were
insufficient to determine a result in that bin.}
\begin{center}
\begin{tabular}{|c|r@{ $\pm$ }r@{ $\pm$ }r|r@{ $\pm$ }r@{ $\pm$ }r|r@{ $\pm$ }r@{ $\pm$ }r|}
\hline
$p_{\rm T}~[{\rm GeV}/c]$ & \multicolumn{3}{c|}{$2.5 < y <3.0$} & \multicolumn{3}{c|}{$3.0 <y<3.5$} & \multicolumn{3}{c|}{$3.5 <y< 4.0$} \\
\hline 
\multicolumn{10}{|c|}{Downstream-track selection} \\
\hline
$0.0-0.2$ & \multicolumn{3}{c|}{---}
                               &   $73$ & $10$ & $2$ &  $40$ &  $8$ &  $1$ \\
$0.2-0.4$ & \multicolumn{3}{c|}{---}
                               &  $278$ & $21$ & $6$ & $288$ & $21$ & $10$ \\
$0.4-0.6$ & $147$ & $15$ & $4$ &  $428$ & $24$ & $7$ & $388$ & $21$ & $10$ \\
$0.6-0.8$ & $202$ & $16$ & $1$ &  $379$ & $22$ & $8$ & $332$ & $21$ &  $8$ \\
$0.8-1.0$ & $176$ & $15$ & $1$ &  $213$ & $16$ & $6$ & $217$ & $17$ &  $1$ \\
$1.0-1.2$ & $113$ & $11$ & $1$ &  $173$ & $14$ & $1$ & $111$ & $12$ &  $4$  \\
$1.2-1.4$ &  $94$ & $11$ & $2$ &   $90$ & $10$ & $0$ &  $32$ &  $8$ &  $0$ \\
$1.4-1.6$ &  $56$ &  $8$ & $2$ &   $64$ &  $8$ & $3$ &  $20$ &  $5$ &  $1$ \\
%
\hline 
\multicolumn{10}{|c|}{Long-track selection} \\
\hline
 $0.0-0.2$ & $17$ & $5$   & $2$   &  $34$ &  $7$   & $3$   & \multicolumn{3}{c|}{---} \\
 $0.2-0.4$ & $31$ & $6$   & $2$   &  $75$ &  $9$   & $4$   & \multicolumn{3}{c|}{---} \\
 $0.4-0.6$ & $63$ & $8$   & $6$   & $121$ & $12$   & $3$   & $41$ & $7$   & $1$   \\
 $0.6-0.8$ & $64$ & $8$   & $2$   & $134$ & $12$   & $3$   & $65$ & $9$   & $5$   \\
 $0.8-1.0$ & $50$ & $7$   & $2$   &  $91$ & $10$   & $2$   & $53$ & $8$   & $4$   \\
 $1.0-1.2$ & $30$ & $6$   & $1$   &  $40$ &  $7$   & $5$   & $35$ & $7$   & $2$   \\
 $1.2-1.4$ & $16$ & $4$   & $0$   &  $33$ &  $6$   & $5$   & $27$ & $5$   & $6$   \\
 $1.4-1.6$ &  $8$ & $3$   & $0$   &  $19$ &  $5$   & $3$   & $14$ & $4$   & $2$   \\
%
\hline
\end{tabular} 
\label{tab:yields}
\end{center}
\end{table}

The beam-gas subtraction and signal yield extraction are repeated for each bin in $p_{\rm T}$ and $y$, 
leading to the results shown in Table~\ref{tab:yields}.
The systematic uncertainties on the extraction of these yields are obtained by 
comparing the yields from single and double Gaussian signal fits 
and from side-band subtraction to the expected yield in a Monte Carlo
sample of comparable statistics to the  data set.
Additionally the fitted and side-band subtracted yields are compared,
and an alternate (exponential) background model is used in the mass fits.
The largest observed deviation in any of these studies is taken as
systematic uncertainty.
For the long-track selection, where the yields 
are lower, the central value 
is obtained from the side-band subtraction method
assuming a linear background.

\section{Efficiency estimation}
\label{sec:efficiency}

A sample of fully simulated events is used 
to estimate the reconstruction and selection
efficiency $\epsilon_i^{\rm sel}$ in each $p_{\rm T}$ and $y$ bin. 
Single $pp$ collisions are generated with the PYTHIA~6.4
program~\cite{PYTHIA6}
and the generated particles are tracked through the detector with
the GEANT~4 package~\cite{GEANT4},
taking into account the details of the
geometry and material composition of the detector.
The simulation of the detector response
is tuned to reproduce test beam results~\cite{LHCb}. In terms of dead and 
noisy channels, the simulation reflects the
detector status of the data set used in this analysis.

Residual misalignments of the tracking stations 
and edge-effects of cell efficiencies in the Outer Tracker 
are not perfectly described in the MC sample, resulting
in an overestimation of the tracking efficiency.
To incorporate these effects, we compare for each detector unit 
the hit content of the tracks
in the data and MC samples 
and randomly remove hits in the simulation until we achieve
agreement 
in all subdetector components and phase-space regions. 
The MC sample modified in this way is the 
nominal MC sample, used throughout the analysis.

To assign systematic uncertainties on the efficiencies obtained in this
MC sample the single track-finding efficiencies were measured. The VELO 
efficiency is obtained by using reconstructed tracks in the TT and in the 
T1--T3 stations and checking for an associated track segment in the VELO. Similarly
the 
TT and T1--T3 station efficiencies are tested by reconstructing tracks using VELO and HCAL information.
For downstream tracks with a $p_{\rm T}$ larger than 0.2~GeV$/c$
agreement between the track-finding efficiencies in data and in the Monte
Carlo sample is observed within the statistical uncertainties of approximately 3\%.
Below 0.2~GeV$/c$, the ratio of efficiencies in data and MC is 
found to be $0.85 \pm 0.12$. As a conservative approach  
3\% (15\%) uncertainties for the reconstruction efficiency of tracks with a $p_{\rm T}$ larger (smaller)
than 0.2~GeV$/c$ are assigned.
Propagating these uncertainties to the \Kshort\ reconstruction efficiency results in
correlated systematic uncertainties of up to 17\% for the 
lowest \Kshort\ $p_{\rm T}$ bins of the downstream-track selection.

The systematic uncertainty on the \Kshort\ selection efficiency is obtained
by comparing, in data and MC, the selection efficiency relative to a preselection. 
This preselection is close to 90\% efficient for downstream-track selected signal events
in MC.

If the reconstruction and selection efficiency varies strongly 
within a given bin of phase space,
the average value estimated with MC will depend on the
assumed production spectrum within the bin.
The extraction of the efficiency-corrected yield in each bin
is therefore repeated using 
efficiencies in four sub-bins rather than an average efficiency,
and the difference with respect to the nominal result is taken
as an uncorrelated systematic uncertainty. The size of this effect 
varies between 0 and 20\%. 
The largest uncertainties are obtained in bins at the limit of the acceptance.

The fraction of non-prompt \Kshort\ signal in the selected MC sample is found to
be 0.6\%.  By definition, this is corrected for in the efficiencies defined in
Eq.~(\ref{eq:epsilon_i_sel}).  Because the correction is so small, even doubling
this contribution would have no significant impact on the final result.
Similarly, the systematic uncertainty due to material interactions, assuming a
conservative $\pm 10\%$ variation of the known detector material, is found
to be negligible.

\begin{figure}
\begin{center}
\epsfig{file=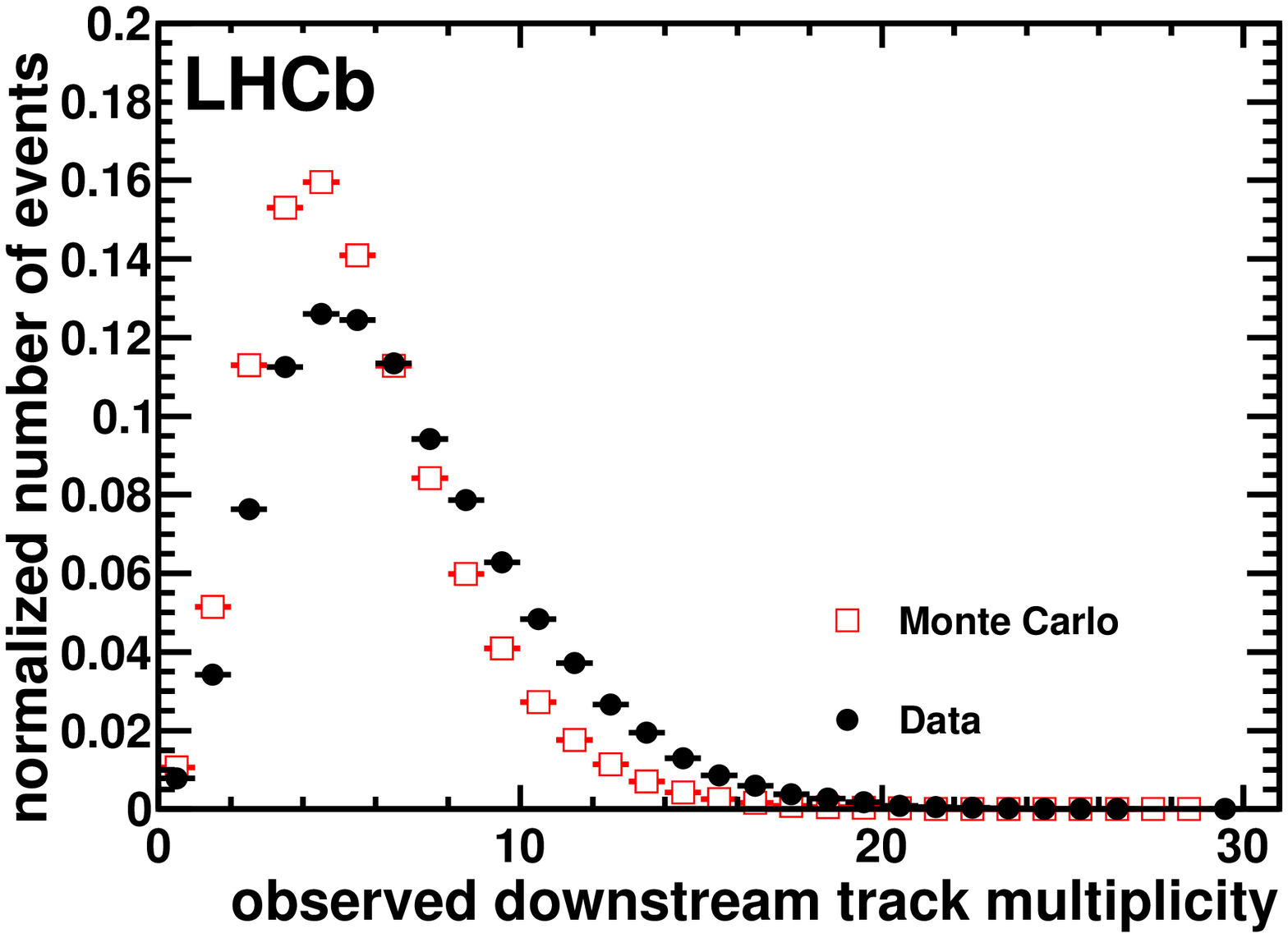,bbllx=10,bblly=10,bburx=553,bbury=390,width=0.49\textwidth}
\epsfig{file=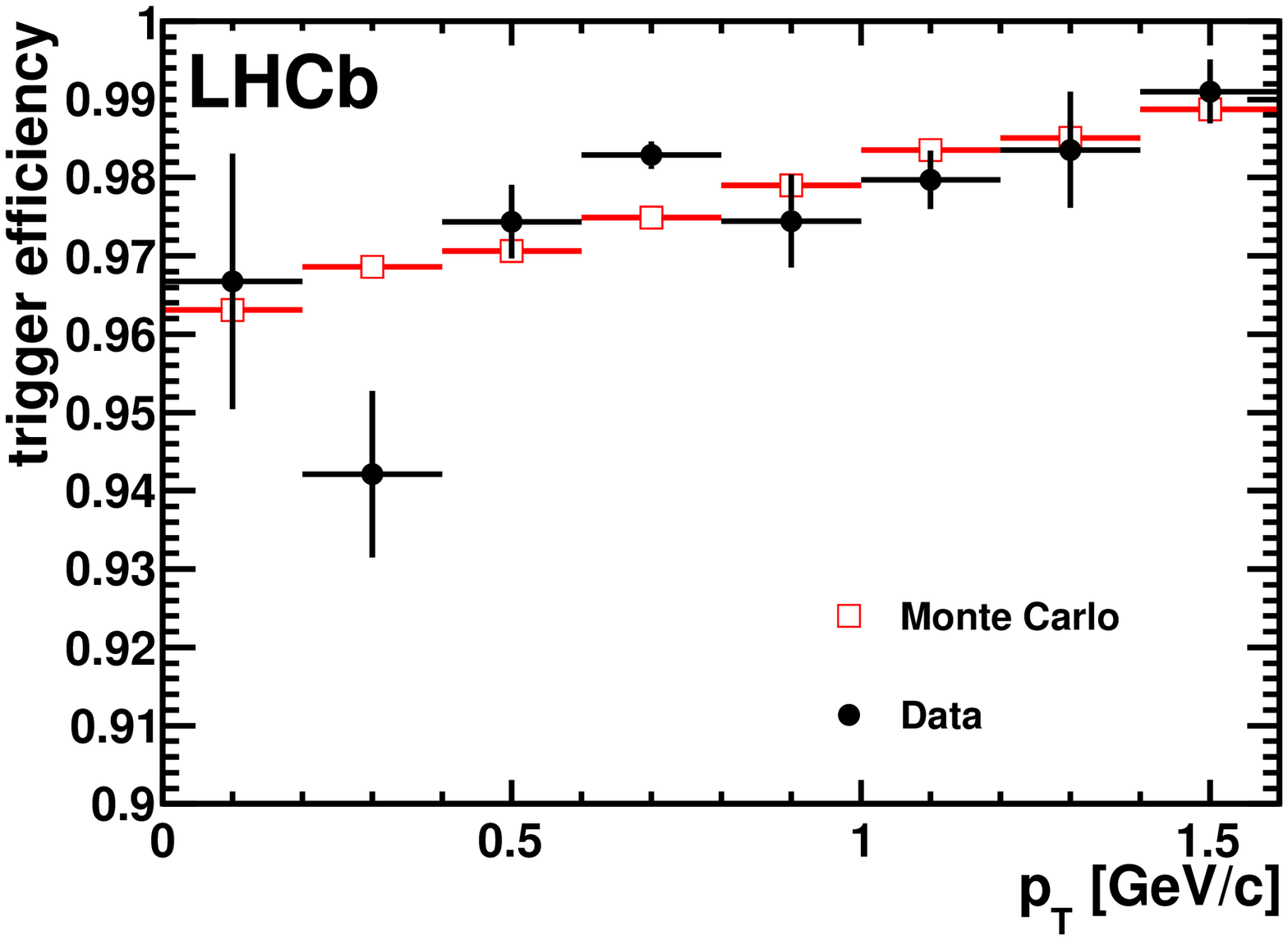,bbllx=10,bblly=10,bburx=553,bbury=390,width=0.49\textwidth}
\caption{\small 
Left:
Downstream track multiplicity for events containing a signal \Kshort,
in data (black filled circles) and MC (red open squares),
normalized to unit area.
Right:
Trigger efficiency for events containing a signal \Kshort\ decay
in the downstream-track selection, 
as a function of the \Kshort\ $p_{\rm T}$, 
estimated both in data (black filled circles) and MC (red open squares),
using Eq.~(\ref{eq:TISTOS}).}
\label{fig:efficiency}
\end{center}
\end{figure}

The trigger efficiency $\epsilon_i^{\rm trig/sel}$ for selected signal events
depends on the track multiplicity. As outlined in Section 3, $\epsilon_i^{\rm
trig/sel}$ is obtained after weighting the previously-defined nominal MC sample
in order to reproduce, in selected signal events, the 
track multiplicity observed in the data (see Fig.~\ref{fig:efficiency} (left)).
This re-weighting is only applied for the determination of the trigger
efficiency, as the reconstruction and selection efficiency has been shown not to
depend on the track multiplicity.  The trigger efficiency is found to be greater
than 95\% in every phase-space bin.  As a cross check, it is also extracted
directly from data,
using a method that exploits the fact that signal events can be triggered by the
\Kshort\ daughters (trigger on signal, TOS) or by the rest of the event (trigger
independent of signal, TIS), with a very large overlap between the two cases.
Assuming that the two ways to trigger are independent, $N_{\rm TIS\&TOS} =
\epsilon_{\rm TIS}~\epsilon_{\rm TOS}~N_{\rm sel} = N_{\rm TIS}~N_{\rm
TOS}/N_{\rm sel}$, where, in a given region of phase space, $N_{\rm TIS}$ and
$N_{\rm TOS}$ are the number of TIS and TOS events, $N_{\rm TIS\&TOS}$ is the
number of events which are simultaneously both TIS and TOS, and $N_{\rm sel}$ is
the number of selected signal events.  Hence
\begin{equation}
\epsilon_{\rm data}^{\rm trig/sel} =
\frac{N_{\rm TIS|TOS}}{N_{\rm sel}} = \frac{N_{\rm TIS|TOS}~N_{\rm
    TIS\&TOS}}{N_{\rm TIS}~ N_{\rm TOS}} \,,
\label{eq:TISTOS}
\end{equation}
where $N_{\rm TIS|TOS}$ is the number of
events which are triggered either as TIS or TOS.
Due to the limited data statistics, a
significant comparison between data and MC can only be done in bands of $p_{\rm
T}$ or $y$, rather than in 2-dimensional bins.  Good agreement is found,
as illustrated in Fig.~\ref{fig:efficiency} (right), and the observed
differences are translated into a global correlated systematic uncertainty of 2\%.

The dependence on the modeling of diffractive processes is studied 
per bin of phase space by changing the fraction of diffractive events 
in the PYTHIA~6.4 sample by 50\% of its value, and by replacing these
events with diffractive events generated with
PYTHIA~8.1~\cite{PYTHIA8}\footnote{%
We consider single- and double-diffractive process types 92--94 in PYTHIA~6.421,
which includes only soft diffraction, and 
103--105 in PYTHIA~8.130 (soft and hard diffraction).
}. 
The evaluation of the MC efficiencies
is repeated for different PYTHIA~6.4 parameter values~\cite{Perugia},
leading to no significant change.

\begin{table}[t] 
\caption{\small Total efficiencies (in \%) 
in bins of transverse momentum $p_{\rm T}$ and rapidity $y$
for the two selections. 
The first uncertainty is uncorrelated, including the statistical 
uncertainty from MC, and the second is at least partially correlated
across bins.}
\begin{center}
\begin{tabular}{|c|r@{ $\pm$ }r@{ $\pm$ }r|r@{ $\pm$ }r@{ $\pm$ }r|r@{ $\pm$ }r@{ $\pm$ }r|}
\hline
$p_{\rm T}~[{\rm GeV}/c]$ & \multicolumn{3}{c|}{$2.5 < y <3.0$} & \multicolumn{3}{c|}{$3.0 <y<3.5$} & 
\multicolumn{3}{c|}{$3.5 <y< 4.0$} \\
\hline
\multicolumn{10}{|c|}{Downstream-track selection} \\
\hline
$0.0-0.2$ & \multicolumn{3}{c|}{---}
                             &  3.4 & 0.5 & 0.5 &  3.0 & 0.6 & 0.3 \\
$0.2-0.4$ & \multicolumn{3}{c|}{---}
                             &  7.3 & 0.2 & 0.8 &  7.4 & 0.2 & 1.0 \\
$0.4-0.6$ &  3.5 & 0.4 & 0.4 & 11.8 & 0.2 & 0.9 & 12.0 & 0.2 & 0.9 \\
$0.6-0.8$ &  7.4 & 0.3 & 0.5 & 15.0 & 0.2 & 1.2 & 15.1 & 0.2 & 1.2 \\
$0.8-1.0$ & 11.1 & 0.2 & 0.9 & 17.1 & 0.2 & 1.3 & 15.8 & 0.4 & 1.2 \\
$1.0-1.2$ & 14.5 & 0.5 & 1.2 & 18.7 & 0.5 & 1.4 & 15.1 & 0.4 & 1.2 \\
$1.2-1.4$ & 16.2 & 0.4 & 1.2 & 18.9 & 0.5 & 1.5 & 13.6 & 1.1 & 1.0 \\
$1.4-1.6$ & 17.8 & 0.6 & 1.3 & 19.1 & 0.7 & 1.5 & 12.6 & 1.2 & 0.9 \\
\hline
\multicolumn{10}{|c|}{Long-track selection} \\
\hline
 $0.0 - 0.2$ &  0.8 &  0.0 &  0.2  &  2.0 &  0.1 &  0.4  &  \multicolumn{3}{c|}{---} \\
 $0.2 - 0.4$ &  0.7 &  0.1 &  0.1  &  2.0 &  0.1 &  0.4  &  \multicolumn{3}{c|}{---} \\
 $0.4 - 0.6$ &  1.2 &  0.0 &  0.2  &  3.7 &  0.1 &  0.6  &  1.3 &  0.3 &  0.2  \\ 
 $0.6 - 0.8$ &  1.9 &  0.1 &  0.3  &  4.9 &  0.1 &  0.6  &  2.9 &  0.1 &  0.4  \\ 
 $0.8 - 1.0$ &  2.6 &  0.1 &  0.3  &  5.6 &  0.1 &  0.7  &  4.1 &  0.5 &  0.5  \\ 
 $1.0 - 1.2$ &  2.8 &  0.1 &  0.3  &  6.1 &  0.5 &  0.6  &  4.3 &  0.3 &  0.4  \\ 
 $1.2 - 1.4$ &  2.7 &  0.2 &  0.2  &  5.7 &  0.5 &  0.5  &  5.1 &  0.6 &  0.6  \\ 
 $1.4 - 1.6$ &  2.8 &  0.3 &  0.2  &  5.7 &  0.5 &  0.5  &  5.4 &  0.5 &  0.5  \\ 
\hline
\end{tabular}  
\label{tab:epsilon}
\end{center} 
\end{table}
 
There are two important differences in the analysis of the \Kshort\ candidates
from the long-track selection relative to the downstream-track selection.
Firstly, a reconstructed PV is required in the former case, so the systematic uncertainty on the 
PV reconstruction efficiency needs to be assessed. The simulation is found to be
in good agreement with the data, but the analysis is more sensitive
to the contribution from diffractive events. Secondly, the background
level in the long-track selection is significantly lower than in the
downstream-track selection, due to the 
PV requirement and the precise VELO measurements.
Therefore it is possible to remove the 
minimum $p_{\rm T}$ requirement on the \Kshort\ daughters in the long-track selection. 
This allows the extension of the analysis to two 
low $p_{\rm T}$ bins in the range $2.5 < y < 3.0$, which are
inaccessible to the downstream-track selection.
The dominant systematic error for these two bins is from the large 
uncertainty on the tracking efficiency for the very 
low $p_{\rm T}$ \Kshort\ daughters.

The estimates of the total efficiencies
$\epsilon_i^{\rm trig/sel}\times\epsilon_i^{\rm sel}$
are given in Table~\ref{tab:epsilon}. The 
various contributions to the uncertainties 
have been classified according to their 
correlations across bins, as shown in Table~\ref{tab:syst}, 
and added in quadrature. 

\section{Results and discussion}
\label{sec:results}

\begin{table}[t] 
\caption{\small Sources of uncertainty on the
\Kshort\ production cross-sections of Eq.~(\ref{eq:sigma_i}), with 
relative values quoted for the downstream-track selection. 
A range of values means that the uncertainty was evaluated per bin of $(p_{\rm T}, y)$ phase space
(with extreme values quoted),
while a single value indicates a global uncertainty assumed to be bin-independent.
The different contributions are classified as uncorrelated
or (at least partially) correlated across the different bins. 
}
\newcommand{\percent}[1]{\ensuremath{#1\,\%~}}
\begin{center}
\begin{tabular}{|l|r|r|r|}
\hline
Source of uncertainty  & \multicolumn{1}{@{\,}c@{\,}|}{uncorrelated} & \multicolumn{1}{c|}{correlated} \\
\hline
{\bf Yields} $N_i^{\rm obs}$  & & \\ 
~ -- Data statistics          & \percent{5-25} &              \\ 
~ -- Signal extraction        & \percent{1-5}  &              \\ 
~ -- Beam-gas subtraction     &                & \percent{<1}   \\ 
\hline
{\bf Efficiency correction}
$(\epsilon_i^{\rm trig/sel}~ \epsilon_i^{\rm sel})^{-1}$ & & \\ 
~ -- MC statistics            & \percent{1-5} &              \\ 
~ -- Track finding            &               & \percent{6-17}\\ 
~ -- Selection                &               & \percent{ 4}  \\ 
~ -- Trigger                  &               & \percent{ 2}\\ 
~ -- $p_{\rm T}$ and $y$ shape within bin & \percent{0-20} & \\ 
~ -- Diffraction modelling    &               & \percent{0-1} \\ 
~ -- Non-prompt contamination &               & \percent{<1}  \\ 
~ -- Material interactions    &               & \percent{<1}  \\ 
\hline
{\bf Normalization} $(L_{\rm int})^{-1}$ & & \\ 
~ -- Bunch currents           &               & \percent{12} \\ 
~ -- Beam widths              &               & \percent{ 5} \\ 
~ -- Beam positions           &               & \percent{ 3} \\ 
~ -- Beam angles              &               & \percent{ 1} \\ 
\hline
Sum in quadrature    & \percent{6-28}& \percent{16-23}\\ 
\hline
\end{tabular}
\label{tab:syst}
\end{center} 
\end{table}
\begin{table}[t] 
\vspace*{-6mm}
\caption{\small Prompt \Kshort\ production cross-section (in $\mu$b) measured in bins of transverse momentum $p_{\rm T}$
and rapidity $y$, as defined in Eq.~(\ref{eq:sigma_i}). 
The first quoted error is the statistical uncertainty, the second error is
the uncorrelated systematic uncertainty, and the third error is 
the systematic uncertainty correlated across bins.}
\begin{center}
\begin{tabular}{|c|r@{~\,$\pm$\,}r@{~\,$\pm$\,}r@{~\,$\pm$\,}r|r@{~\,$\pm$\,}r@{~\,$\pm$\,}r@{~\,$\pm$\,}r|r@{~\,$\pm$\,}r@{~\,$\pm$\,}r@{~\,$\pm$\,}r|}
\hline
$p_{\rm T}~[{\rm GeV}/c]$ & \multicolumn{4}{c|}{$2.5 < y <3.0$} & \multicolumn{4}{c|}{$3.0 <y<3.5$} & 
\multicolumn{4}{c|}{$3.5 <y< 4.0$} \\
\hline
 $0.0 - 0.2$ &  294 &   80 &   38 &   90 &  316 &   43 &   44 &   72 &  196 &   39 &   39 &   38 \\ 
 $0.2 - 0.4$ &  649 &  133 &  136 &  183 &  562 &   42 &   22 &  101 &  571 &   42 &   25 &  114 \\ 
 $0.4 - 0.6$ &  618 &   63 &   66 &   97 &  534 &   30 &   12 &   86 &  477 &   26 &   14 &   77 \\ 
 $0.6 - 0.8$ &  401 &   32 &   18 &   64 &  371 &   21 &    9 &   59 &  323 &   20 &    9 &   51 \\ 
 $0.8 - 1.0$ &  232 &   20 &    4 &   37 &  183 &   14 &    6 &   29 &  201 &   16 &    6 &   33 \\ 
 $1.0 - 1.2$ &  115 &   11 &    4 &   18 &  136 &   11 &    3 &   22 &  108 &   12 &    5 &   17 \\ 
 $1.2 - 1.4$ &   85 &   10 &    3 &   14 &   70 &    8 &    2 &   11 &   35 &    9 &    3 &    6 \\ 
 $1.4 - 1.6$ &   46 &    7 &    2 &    7 &   49 &    6 &    3 &    8 &   23 &    6 &    2 &    4 \\ 
\hline
\end{tabular}  
\label{tab:results}
\end{center} 
\end{table}
 
The cross-sections defined in Eq.~(\ref{eq:sigma_i}) are
evaluated separately for both the downstream- and long-track selections. 
In every phase-space bin, the two sets of results are found to be consistent with each other.
The relative uncertainties
on the measurement for the downstream-track selection are summarized in Table~\ref{tab:syst}.
Since the downstream- and long-track results are not statistically independent, and since the 
downstream-track selection contains already most of the statistical power in bins where
a measurement is possible, the measurements are not combined. The final results, listed in Table~\ref{tab:results}, 
are taken from the downstream-track selection, 
except in the two lowest $p_{\rm T}$ bins for $2.5<y<3.0$ where they are taken from the 
long-track selection. 

\begin{figure}[t]
\vspace*{-10mm}
\begin{center}
\epsfig{file=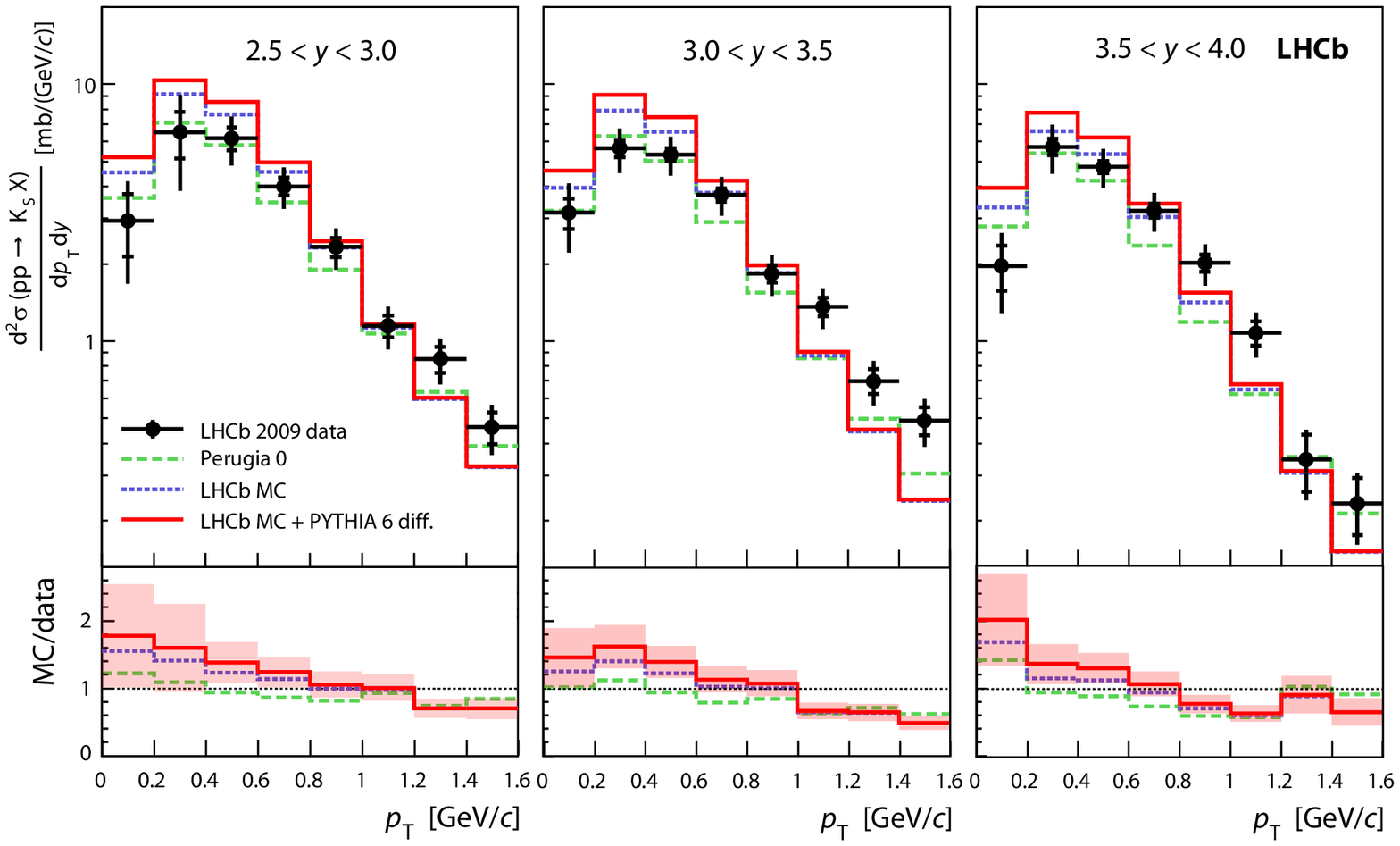,bbllx=10,bblly=5,bburx=555,bbury=330,width=1.0\textwidth}
\caption{\small Double-differential prompt \Kshort\ production cross-section in
$pp$ collisions at $\sqrt{s}=0.9$~TeV as a function of transverse momentum
$p_{\rm T}$ and rapidity $y$.  The points represent LHCb data, with total
uncertainties shown as vertical error bars and statistical uncertainties as tick
marks on the bars.  The histograms are predictions from different
settings of the PYTHIA generator (see text).  The lower plots show the MC/data
ratios, with the shaded band representing the uncertainty for one of these
ratios, dominated by the uncertainty on the measurements (the relative
uncertainties for the other ratios are similar).}
\label{fig:spectra}
\end{center}
\end{figure}

The corresponding differential cross-sections are shown in Fig.~\ref{fig:spectra} as function of transverse
momentum for the three different rapidity bins. They 
include both non-diffractive and diffractive prompt \Kshort\ production,
and are compared with three different sets of predictions, all obtained with 
the PYTHIA~6.4 generator~\cite{PYTHIA6}.
These predictions are represented as histograms in Fig.~\ref{fig:spectra} and
correspond to:
\begin{itemize}
\item the LHCb settings\footnote{%
We use PYTHIA~6.421, and include 
process types 11--13, 28, 53, 68, 91--95, 421--439, 
461--479
with non-default parameter values 
ckin(41)\!\! =3.0,
mstp(2)\!\! =2,
mstp(33)\!\! =3,
mstp(128)\!\! =2,
mstp(81)\!\! =21,
mstp(82)\!\! =3,
mstp(52)\!\! =2,
mstp(51)\!\! =10042,
parp(67)\!\! =1.0,
parp(82)\!\! =4.28,
parp(89)\!\! =14000,
parp(90)\!\! =0.238,
parp(85)\!\! =0.33,
parp(86)\!\! =0.66,
parp(91)\!\! =1.0,
parp(149)\!\! =0.02,
parp(150)\!\! =0.085,
parj(11)\!\! =0.5,
parj(12)\!\! =0.4,
parj(13)\!\! =0.79,
parj(14)\!\! =0.0,
parj(15)\!\! =0.018,
parj(16)\!\! =0.054,
parj(17)\!\! =0.131,
mstj(26)\!\! =0,
parj(33)\!\! =0.4.
The particle decay probabilities are computed using EvtGen \cite{EVTGEN}. 
}, 
which include only soft diffraction as described by PYTHIA~6.4 (red solid histogram);
\item the LHCb settings where diffractive processes have been switched off (blue dotted histogram);
\item the ``Perugia 0'' settings~\cite{Perugia}, which exclude diffraction 
(green dashed histogram).
\end{itemize} 
The predictions agree reasonably well with the data, although they 
tend to underestimate (overestimate) the measured production
in the highest (lowest) $p_{\rm T}$ bins. 

\begin{figure}[t]
\vspace*{-10mm}
\begin{center}
\epsfig{file=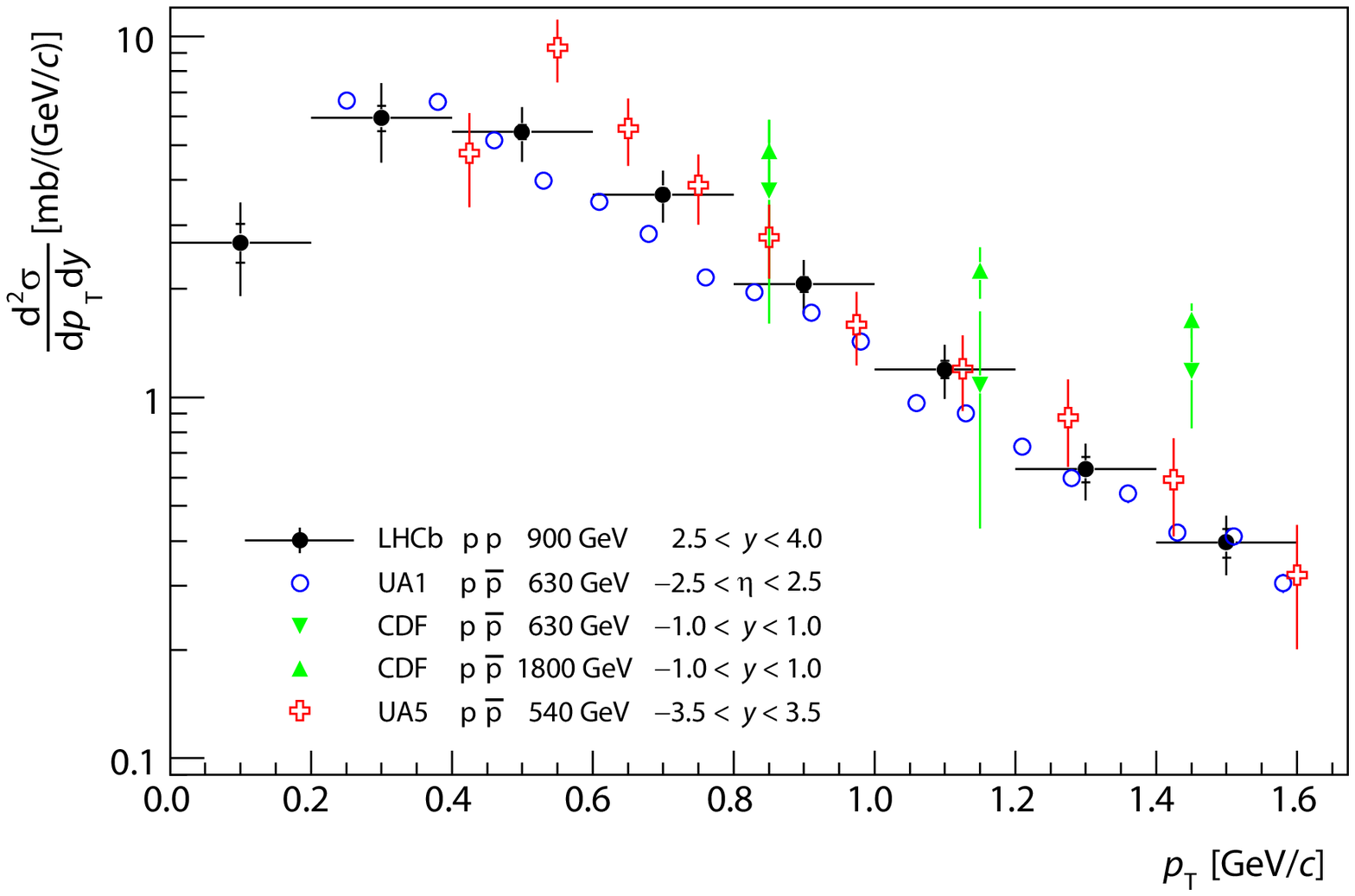,bbllx=3,bblly=10,bburx=515,bbury=348,width=0.85\textwidth}
\vspace*{-4mm} 
\caption{\small Absolute measurements of the prompt \Kshort\ production
cross-section as a function of transverse momentum $p_{\rm T}$, performed
by the UA1~\cite{UA1KS}, UA5~\cite{UA5KS2}, CDF~\cite{CDFKSabs} and LHCb experiments, 
at different high-energy hadron colliders
and in different rapidity ($y$) or pseudo-rapidity ($\eta$) ranges.}
\label{fig:compare}
\end{center}
\end{figure}

Previous measurements of the prompt \Kshort\ cross-section
in high-energy $p\bar{p}$ collisions,
performed by UA5~\cite{UA5KS2}, UA1~\cite{UA1KS} and CDF~\cite{CDFKSabs}
at different centre-of-mass energies and in different rapidity or 
pseudo-rapidity regions, have been published in the form of 
invariant differential cross-sections 
$E ~ {\rm d}^3\sigma/{\rm d}^3p$ as a function of $p_{\rm T}$. We convert these into 
measurements of $d^2\sigma/({\rm d}p_{\rm T}{\rm d}y)$ by multiplication with $2\pi p_{\rm T}$, 
and compare them with our results in Fig.~\ref{fig:compare}, limiting the $p_{\rm T}$
range of previous measurements to $1.6~{\rm GeV}/c$. In this figure, LHCb results are shown for the 
rapidity range $2.5<y<4.0$, obtained by averaging the results for the three separate $y$ bins,
assuming conservatively that the correlated systematic uncertainties are 100\% correlated.
In general the agreement is reasonable, given the
spread of centre-of-mass energies and the fact that the results are averaged
over different ranges in rapidity or pseudo-rapidity. 
The ability of LHCb to contribute measurements that
extend the kinematic range towards high rapidities and very low $p_{\rm T}$ is apparent. 

\section{Conclusions}
\label{sec:conclusion}

Studies of prompt \Kshort\ production at $\sqrt{s}=0.9$~TeV have been presented,
made with the LHCb detector using the first $pp$ collisions delivered by the LHC
during 2009.  The data sample used corresponds to an integrated luminosity of
$6.8 \pm 1.0~\mu{\rm b}^{-1}$, a value which has been determined using
measurements of the beam profiles that exploit the high precision of the VELO.
This is the most precise determination of the luminosity for the 2009 LHC pilot
run, only limited by the uncertainties on the beam intensity.

The differential cross-section
has been measured as a function of $p_{\rm T}$ and $y$, over a range
extending down to $p_{\rm T}$ less than 0.2~GeV$/c$, and in the rapidity
interval $2.5 < y < 4.0$, a region that has not been explored
in previous experiments at this energy. 
These results show reasonable consistency
with expectations based on the PYTHIA 6.4 generator, and should provide 
valuable input for the future tuning of Monte Carlo generators. 

\section*{Acknowledgments}
\label{sec:acknowledgments}
\currentpdfbookmark{Acknowledgments}{acknowledgments}

We express our gratitude to our colleagues in the CERN accelerator departments for the excellent performance of the LHC.
The valuable contribution of J.-J.~Gras (CERN) to the analysis of the LHC beam
current measurements is gratefully acknowledged.
We thank the technical and administrative staff at CERN and at the LHCb institutes, and acknowledge support from the National Agencies: CAPES, CNPq, FAPERJ and FINEP (Brazil); CERN; NSFC (China); CNRS/IN2P3 (France); BMBF, DFG, HGF and MPG (Germany); SFI (Ireland); INFN (Italy); FOM and NWO (Netherlands); SCSR (Poland); ANCS (Romania); MinES of Russia and Rosatom (Russia); MICINN, XUNGAL and GENCAT (Spain); SNSF and SER (Switzerland); NAS Ukraine (Ukraine); STFC (United Kingdom); NSF (USA).
We also acknowledge the support received from the ERC under FP7 and the
R\'{e}gion Auvergne.

\raggedright

\end{document}